\documentclass[12pt]{article}
\usepackage{cite}
\usepackage{enumerate}
\usepackage{ifpdf}
\usepackage{ae}
\usepackage[T1]{fontenc}
\usepackage[ansinew]{inputenc}
\usepackage{mathrsfs}
\usepackage{amsmath}
\usepackage{relsize}
\usepackage{amssymb}
\usepackage{tikz}
\usepackage{graphicx}
\usepackage{color}
\definecolor{darkblue}{cmyk}{0.9,0.9,0,0}
\definecolor{darkgreen}{rgb}{0,0.55,0}
\usepackage[colorlinks=true,linkcolor=darkblue,citecolor=darkblue,urlcolor=darkblue]{hyperref}
\usepackage{epsfig}
\usepackage{epstopdf}
\usepackage{graphicx}

\makeatletter
\long\def\@makecaption#1#2{
  \vskip\abovecaptionskip
  \sbox\@tempboxa{{\captionfonts #1: #2}}
  \ifdim \wd\@tempboxa >\hsize
    {\captionfonts #1: #2\par}
  \else
    \hbox to\hsize{\hfil\box\@tempboxa\hfil}
  \fi
  \vskip\belowcaptionskip}
\makeatother

\newcommand{\beq}{\begin{equation}}
\newcommand{\eeq}{\end{equation}}
\newcommand{\beqy} {\begin{eqnarray}}
\newcommand{\eeqy} {\end{eqnarray}}
\newcommand{\bsmat}{\begin{smallmatrix}}
\newcommand{\esmat}{\end{smallmatrix}}
\newcommand{\bmat}{\begin{matrix}}
\newcommand{\emat}{\end{matrix}}
\def\zb{{\bar{z}}}
\def\Li{\textrm{Li}}

\def\({\left(}
\def\){\right)}
\def\[{\left[}
\def\]{\right]}

\def\<{\langle}
\def\>{\rangle}

\usepackage{varioref}
\usepackage{makeidx}
\makeindex

\usepackage[english]{babel}
\usepackage{caption}
\usepackage{subfig}

\usepackage{tabularx}
\begin{document}

\thispagestyle{empty}

\renewcommand{\thefootnote}{\fnsymbol{footnote}}
\setcounter{page}{1}
\setcounter{footnote}{0}
\setcounter{figure}{0}
\begin{titlepage}

\begin{center}

\vskip 2.3 cm 

\vskip 5mm

{\Large \bf
On Genus-one String Amplitudes on $AdS_5 \times S^5$
}
\vskip 0.5cm

\vskip 15mm

\centerline{Luis F. Alday}
\bigskip
\centerline{\it Mathematical Institute, University of Oxford,} 
\centerline{\it Woodstock Road, Oxford, OX2 6GG, UK}

\end{center}

\vskip 2 cm

\begin{abstract}
\noindent 
We study non-planar correlators in ${\cal N}=4$ super-Yang-Mills in Mellin space. We focus in the stress tensor four-point correlator to order $1/N^4$ and in a strong coupling expansion. This can be regarded as the genus-one four-point graviton amplitude of type IIB string theory on $AdS_5 \times S^5$ in a low energy expansion. Both the loop supergravity result as well as the tower of stringy corrections have a remarkable simple structure in Mellin space, making manifest important properties such as the correct flat space limit and the structure of UV divergences. 

\end{abstract}
\end{titlepage}


\setcounter{page}{1}
\renewcommand{\thefootnote}{\arabic{footnote}}
\setcounter{footnote}{0}

 \def\nref#1{{(\ref{#1})}}
 
\section{Introduction}
Scattering amplitudes are one of the most fundamental observables in a quantum field theory. Over the last decades remarkable mathematical structures underlying gauge and gravity amplitudes have been discovered. Most of them not apparent from a Lagrangian formulation of the theory. The study of the ultraviolet (UV) structure of gravity theories through graviton scattering has a long history. While pure gravity in four dimensions, described by the Einstein-Hilbert action, is finite at one loop, explicitly computations show that UV divergences appear at two loops \cite{Goroff:1985th,vandeVen:1991gw}. In supersymmetric versions of gravity cancelations between bosonic and fermionic contributions delay the loop order at which UV divergences can appear. Moreover, hidden symmetries can lead to finite results at orders where UV divergences are naively expected. Maximally supersymmetric theories of gravity have been analysed in various dimensions. In the absence of new unknown cancelation mechanisms, it is believed that the maximally supersymmetric theory in four dimensions, ${\cal N}=8$ supergravity, will be UV divergent at seven loops, see for instance \cite{Vanhove:2009zz}. While an explicit computation to seven loops is still out of reach, and the presence of UV divergences still a matter of debate, the computation of \cite{Bern:2018jmv} shows that UV divergences are present at five loops for $D=24/5$. It has been argued that any new mechanisms in four dimensions at seven loops, should have appeared for five loops at  $D=24/5$. This would suggest  ${\cal N}=8$ supergravity is not finite, even perturbatively. 

A mechanism to cure UV divergences is provided by string theory, where point particles are replaced by strings of finite size $\sqrt{\alpha'}$. String theory provides an ultraviolet completion of (super)gravity. Its low energy dynamics is described in terms of an effective action, which contains a super-symmetric version of the Einstein Hilbert action, plus an infinite tower of higher derivative terms (stringy corrections) weighted by powers of $\alpha'$. The structure of this low energy effective action can be inferred from string scattering amplitudes. Furthermore, by studying the scattering of the graviton state in string theory we can learn much not only about the structure of string theory, but also about the maximal supergravity theories that string theory UV completes. Over the last decade there has been great progress studying the four-supergraviton amplitude in type II string theory in a low energy expansion in flat space, see for instance \cite{Green:2008uj,Green:2010sp,DHoker:2014oxd,DHoker:2015gmr}. The four-supergraviton amplitude depends on the string coupling constant $g_s$, the string size $\alpha'$ and the momenta and helicities of the external graviton states. The dependence on the helicities is captured through the Lorentz scalar ${\cal R}^4$, which enters as a prefactor and will be suppressed. The dependence on the momenta is through the Mandelstam variables $s,t,u$, with $s+t+u=0$. In string perturbation theory the amplitude admits an expansion in powers of $g_s$, where the power $g_s^{2h-2}$ corresponds to the contribution from genus $h$ worldsheets:
\begin{equation}
A(g_s,\alpha',s,t) = \frac{1}{g_s^2} A^{\text{tree}}(\alpha',s,t) + A^{\text{loop}}(\alpha',s,t) + g_s^2 A^{\text{2-loop}}(\alpha',s,t) +\cdots
\end{equation}
Even in flat space explicit results are only known up to genus two, {\it i.e.} two loops. The tree level result is given by the Virasoro-Shapiro amplitude \cite{Virasoro:1969me}:
\begin{equation}
A^{\text{tree}}(\alpha',s,t) = \frac{\Gamma(-\alpha' s/4)\Gamma(-\alpha' t/4)\Gamma(-\alpha' u/4)}{\Gamma(1+\alpha' s/4)\Gamma(1+\alpha' t/4)\Gamma(1+\alpha' u/4)}
\end{equation}
The one loop contribution was computed in \cite{Green:1981yb} as an integral expression, and its low energy expansion was studied in \cite{Green:2008uj}. The two loop contribution was computed in \cite{DHoker:2001kkt,DHoker:2005vch}, see also \cite{Berkovits:2005ng}, and its low energy expansion was studied in \cite{DHoker:2017pvk}. Although these results are quite complicated, much can be learnt from them. 

In this paper we would like to tackle the problem of computing graviton string amplitudes in $AdS_5 \times S^5$ to genus one and in a low energy expansion. At present we don't have a systematic way to directly compute string amplitudes in curved space-time. However, for the special case of $AdS_5 \times S^5$ the AdS/CFT duality offers an alternative approach. In a compact space one cannot define asymptotic states but the $AdS/CFT$ duality dictates that one should map string amplitudes in the bulk to correlators of local operators in the boundary. The $AdS/CFT$ duality relates type IIB string theory on $AdS_5 \times S^5$ to four-dimensional ${\cal N}=4$ SYM with the following identifications
\begin{equation}
\label{AdS/CFT}
g^2_s \sim \frac{\lambda^2}{c},~~~~\frac{R^2}{\alpha'} = \sqrt{\lambda}
\end{equation}
where $c=(N^2-1)/4$ is the central charge of the gauge theory and $\lambda=g_{YM}^2N$ is the t' Hooft coupling. The graviton on $AdS$ is dual to the stress tensor super multiplet in ${\cal N}=4$ SYM, and different members of the multiplet correspond to different helicities of the graviton. The superconformal primary of the stress tensor multiplet is denoted by ${\cal O}_2$, a scalar operator of protected dimension two. In this paper we will consider the four-point correlator of such operators $\langle {\cal O}_2{\cal O}_2{\cal O}_2{\cal O}_2 \rangle \sim {\cal G}(U,V)$, where $U$ and $V$ are the usual conformal cross-ratios. More precisely, it has been argued \cite{Penedones:2010ue,Fitzpatrick:2011ia,Paulos:2011ie} that the appropriate quantity to associate to a scattering amplitude in AdS is the Mellin transform of the above correlator, roughly given by
\begin{equation}
{\cal G}(U,V) \sim \int_{-i \infty}^{i \infty} ds dt U^{s/2} V^{t/2} M(s,t).
\end{equation}
Perhaps the strongest evidence to support this claim is that in the flat space limit, where the radius of $AdS$ becomes very large, and the Mellin variables are rescaled accordingly, one recovers scattering amplitudes in flat space. According to the identification (\ref{AdS/CFT}) the loop/genus expansion of the AdS amplitude corresponds to the large central charge expansion of the correlator. Furthermore the low energy expansion corresponds to an expansion around large $\lambda$, where the leading term corresponds to the supergravity result and the tower of terms suppressed by powers of $1/\lambda$ corresponds to stringy corrections. 

The study of gravitational theories on AdS by CFT correlators in a large $c$ expansion was initiated in \cite{Heemskerk:2009pn} at tree level. The technology developed in \cite{Alday:2016njk,Aharony:2016dwx} allowed to push this program further, giving access to loop amplitudes on AdS. In a very precise sense the one loop result ${\cal G}^{\text{loop}}(U,V)$ follows from the square of the tree-level result ${\cal G}^{\text{tree}}(U,V)$, following an AdS unitarity method. These ideas were applied to the specific correlator at hand in \cite{Alday:2017xua}, where loop corrections to the supergravity result, without stringy corrections, were computed. See \cite{Aprile:2017bgs} for an alternative approach to the same problem. An important feature of these computations is the presence of operator mixing. To solve the mixing problem one needs to consider more general correlators $\langle {\cal O}_2{\cal O}_2{\cal O}_p{\cal O}_p \rangle$ at tree level, where ${\cal O}_p$ is a tower of protected scalar operators of dimension $\Delta=p$, which corresponds to the KK-modes of the graviton. From the AdS perspective this has the following interpretation: even if we consider gravitons as the external states, KK-modes will run along the loop, see figure.
\begin{figure}[h]
\includegraphics[scale=0.32]{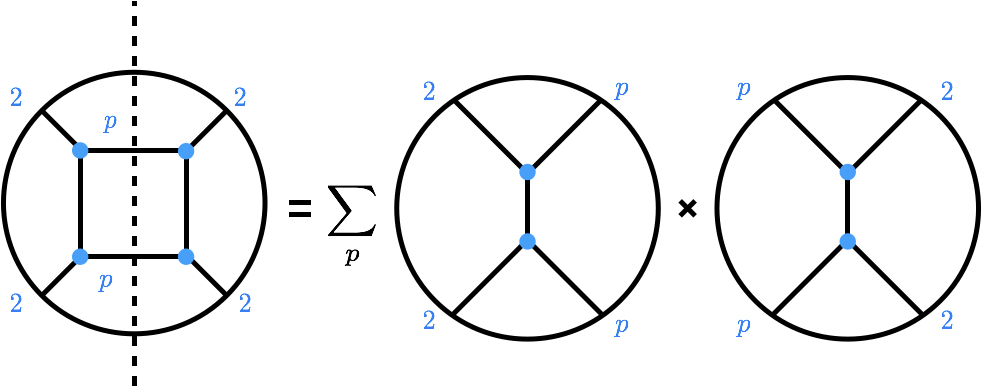}
\centering
\label{fig:box22pp}
\end{figure}
In \cite{Alday:2018pdi} a machinery to compute stringy corrections to the above loop supergravity result was developed. It was shown that in the flat space limit the results are in perfect agreement with the low energy expansion of the genus one string amplitude considered in \cite{Green:2008uj}. The computation was done at the level of the space-time answer ${\cal G}^{\text{loop}}(U,V)$ and focusing in certain piece of the answer, namely the double-discontinuity, that contains in principle all physical information. 

In the present note we study the structure of the loop answer in Mellin space. We will consider both, the loop supergravity contribution as well as the stringy corrections. In order to compute the latter we will assume the Mellin amplitude at tree level, for all correlators $\langle {\cal O}_2{\cal O}_2{\cal O}_p{\cal O}_p \rangle$, is given. We will see that the structure of both contributions in Mellin space is remarkably simple. The result in Mellin space is illuminating in several respects. First, it allows a direct comparison with the corresponding amplitude in flat space. In curved space time the result involves an infinite sum over poles, which reduces to branch cuts in the corresponding flat space limit. Second, the structure and degree of UV divergences is much more transparent, since they arise from 'divergent' sums in the definition of the Mellin amplitude. These sums can be regularised, at the expense of adding polynomial ambiguities. At each order in $\alpha'$, however, the degree of such ambiguities is fixed by requiring analyticity in the spin in the region where the inversion integral of \cite{Caron-Huot:2017vep} converges. As a byproduct of our studies we also derive an integral formula that computes the anomalous dimensions of double trace operators for any polynomial Mellin amplitude. This can be useful in different contexts. 

This paper is organised as follows. In the next section we describe the structure in Mellin space of the supergravity result, together with the whole tower of stringy corrections. Then in section \ref{loopstructure} we describe the structure of the genus one result, including both, the loop supergravity result, as well as the infinite tower of stringy corrections. The loop supergravity result is given in full detail, while for the stringy corrections we give an algorithm to compute them, from the result at tree level. Our main tool is a basis of polynomial functions in Mellin space, described in the appendices, with a prescribed anomalous dimension. At the end of the section we compare our results to the corresponding results in flat space. Then we conclude with a list of open problems. Several appendices are also included.  In particular, we describe an algorithm to efficiently compute the anomalous dimension due to any polynomial Mellin amplitude. This could be useful in other contexts. 
 
\section{Geralities and structure at tree level}
\subsection{Generalities}

${\cal N}=4$ SYM possesses a tower of half-BPS operators ${\cal O}_p$, with $p=2,3,\cdots$, of dimension $\Delta=p$ and transforming in the $[0,p,0]$ representation of the $SU(4)$ $R-$symmetry group. These operators map to the KK-modes of the graviton on $AdS_5 \times S^5$. It is convenient to represent them as follows
\begin{equation}
{\cal O}_p(x,y) = y^{I_1} \cdots y^{I_p}  Tr\left(\varphi^{I_1}(x) \cdots \varphi^{I_p}(x)\right), 
\end{equation}
where the $SU(4)$ indices have been contracted with a null vector $y^I, I=1,\cdots,6$. We will consider a special class of correlators involving these operators
 \begin{equation}
 \langle {\cal O}_2(x_1,y_1){\cal O}_2(x_2,y_2){\cal O}_p(x_3,y_3){\cal O}_p(x_4,y_4)\rangle = \frac{(y_1 \cdot y_2)^2 (y_3 \cdot y_4)^p }{x_{12}^4 x_{34}^{2p}} {\cal G}_p(U,V,\alpha,\bar \alpha) 
 \end{equation}
 where we have introduced space-time and $R-$symmetry cross-ratios 
 \begin{eqnarray}
U&=& \frac{x_{12}^2 x_{34}^2}{x_{13}^2 x_{24}^2} ,~~~ V= \frac{x_{23}^2 x_{14}^2}{x_{13}^2 x_{24}^2},\\
\alpha \bar \alpha&=& \frac{y_{12}^2 y_{34}^2}{y_{13}^2 y_{24}^2},~~~ (1-\alpha)(1-\bar \alpha)= \frac{y_{23}^2 y_{14}^2}{y_{13}^2 y_{24}^2},
 \end{eqnarray}
The correlator ${\cal G}_p(U,V, \alpha, \bar \alpha)$ can be split into the contribution from short, protected, multiplets and long multiplets
\begin{equation}
{\cal G}_p(U,V,\sigma,\tau)  = {\cal G}^{\text{short}}_p(U,V,\alpha, \bar \alpha) + {\cal G}^{\text{long}}_p(U,V,\alpha, \bar \alpha) 
\end{equation}
${\cal G}^{\text{short}}_p(U,V,\alpha, \bar \alpha)$ does not depend on the coupling constant and can be computed following \cite{Beem:2016wfs}. For this class of correlators superconformal Ward identities fix completely the R-charge dependence of the long contribution, see \cite{Eden:2000bk,Nirschl:2004pa,Rastelli:2017udc} 
 \begin{eqnarray}
 {\cal G}^{\text{long}}_p(U,V,\alpha, \bar \alpha) = \frac{(z-\alpha)(z-\bar{\alpha})(\bar z - \alpha)(\bar z -\bar \alpha)}{\alpha^2 \bar \alpha^2}{\cal H}_p(U,V)
 \end{eqnarray}
where we have introduced $U= z \bar z,V=(1-z)(1-\bar z)$. ${\cal H}_p(U,V)$ encodes the dynamically non-trivial information in the correlator and admits a decomposition in super conformal blocks
\begin{equation}
{\cal H}_p(U,V) = \sum_{\Delta,\ell} a_{\Delta,\ell} g_{\Delta,\ell}(z,z)
\end{equation}
where the sum runs over superconformal primaries in long multiplets, of dimension $\Delta$ and spin $\ell$, present in the OPEs ${\cal O}_2 \times {\cal O}_2$ and ${\cal O}_p \times {\cal O}_p$, transforming in the singlet of $SU(4)$. $a_{\Delta,\ell}$ denote the OPE coefficients with which such primaries appear. The explicit expression for the superconformal blocks in given in appendix \ref{basis}. We will study this correlator in a double expansion. First around large central charge
 \begin{equation}
{\cal H}_p(U,V) = {\cal H}^{\text{MFT}}_p(U,V) + \frac{1}{c} {\cal H}^{\text{tree}}_p(U,V) + \frac{1}{c^2} {\cal H}^{\text{loop}}_p(U,V)  + \cdots
 \end{equation}
and then each term around large 't Hooft coupling $\lambda$.  In this regime the intermediate operators are double trace operators labelled by their spin and $n=0,1,\cdots$, with dimension
\begin{equation}
\Delta_{n,\ell} = 4+ 2n + \ell + \frac{1}{c}\gamma^{\text{tree}}_{n,\ell}+ \frac{1}{c^2}\gamma^{\text{loop}}_{n,\ell} + \cdots
\end{equation}
and OPE coefficients denoted by $a_{p,n,\ell}$:
\begin{equation}
a_{p,n,\ell} = a^{\text{MFT}}_{p,n,\ell} +\frac{1}{c} a^{\text{tree}}_{p,n,\ell} + \frac{1}{c^2}a^{\text{loop}}_{p,n,\ell}  + \cdots
\end{equation}
where the OPE coefficients in the {\it mean field theory} approximation follow from ${\cal H}^{\text{MFT}}_p(U,V)$ and are given by $a^{\text{MFT}}_{p,n,\ell}=c^{\text{MFT}}_{p,n,\ell} c^{\text{MFT}}_{2,n,\ell}$ with
\begin{eqnarray}
\label{MFTOPE}
(c^{\text{MFT}}_{p,n,\ell})^2 &=& \frac{24 (\ell+1) \Gamma(n+1) (\ell+2 n+6) \Gamma^2 (n+3) \Gamma (\ell+n+2) \Gamma^2 (\ell+n+4) }{p^2 (p+1) \Gamma (n+5) \Gamma (2 n+5) \Gamma (p-1) \Gamma^3 (p) \Gamma (\ell+n+6) \Gamma (2 \ell+2 n+7)} \nonumber \\
& &\times \frac{\Gamma (n+p+3) \Gamma (\ell+n+p+4)}{ \Gamma (n-p+3) \Gamma (\ell+n-p+4)}
\end{eqnarray}
In this paper we will study the structure of the $1/c$ corrections to the correlator in Mellin space, defined as
\begin{equation}
{\cal H}_p(U,V) = \int_{-i \infty}
^{i \infty} \frac{ds dt}{(4\pi i)^2} U^{\frac{s}{2}} V^{\frac{t-p-2}{2}} {\cal M}_p(s,t) \Gamma\left(\tfrac{2p-s}{2} \right) \Gamma\left(\tfrac{4-s}{2} \right) \Gamma\left(\tfrac{p+2-t}{2} \right)^2 \Gamma\left(\tfrac{p+2-u}{2} \right)^2
\end{equation}
with $s+t+u=2p$. As a consequence of crossing symmetry
\begin{equation}
\label{sym}
{\cal M}_p(s,t) = {\cal M}_p(s,u),~~~~{\cal M}_2(s,t)={\cal M}_2(t,s)
\end{equation}
When referring to the particular case $p=2$, the index $2$ will often be suppressed, hence  ${\cal M}(s,t) \equiv {\cal M}_2(s,t)$. Let us start by describing the tree level result. 

\subsection{Structure at tree level}
The Mellin amplitude at tree level ${\cal M}^{\text{tree}}_p(s,t)$ includes the supergravity result plus an infinite tower of stringy corrections, suppressed by powers of $1/\lambda$. The supergravity solution takes the form 
 \begin{equation}
 {\cal M}^{\text{tree-sugra}}_p(s,t) = \frac{4p}{\Gamma(p-1)} \frac{1}{(s-2)(t-p)(u-p)}.
 \end{equation}
Stringy corrections arise as 'truncated' solutions to the crossing equations, where only operators with finite support in the spin acquire a correction, see \cite{Heemskerk:2009pn}. They correspond to quartic bulk vertices of the schematic form $\phi_2^2 \nabla^m \phi_p^2$, where $\phi_p$ are the KK scalars dual to ${\cal O}_p$. In Mellin space, they are given by polynomials in the Mellin variables with the symmetry properties (\ref{sym}), see \cite{Penedones:2010ue,Alday:2014tsa}. We will denote such polynomials by $V^{(q)}_{p}(s,t)$. For each vertex the support in the spin is given by the degree in the variable $t$ (which is always even). The full tree level amplitude then takes the form
 \begin{equation}
 {\cal M}^{\text{tree}}_p(s,t) = \frac{p}{\Gamma(p-1)}\left( \frac{4}{(s-2)(t-p)(u-p)}  + \sum_{q} \lambda^{-3/2-d(q)/2} V^{(q)}_{p}(s,t) \right)
 \end{equation}
Based in consistency with the flat space limit, both at tree level and at one loop, it has been argued in \cite{Alday:2018pdi}, that each vertex $V^{(q)}_{p}(s,t)$ is also a polynomial in $p$. Finally each vertex is suppressed by a power $\lambda^{-3/2-d(q)/2}$ where $d(q)$ is the total degree of the polynomial in the Mellin variables. For instance, the first vertices are given by
\begin{eqnarray}
V^{(0)}_p(p,s,t) &=& \zeta_3 (p+1)_3\\
V^{(1)}_p(p,s,t) &=& \frac{\zeta_5}{8} (p+1)_5 \left( s^2+t^2+u^2 + \frac{2p(p-2)}{(p+5)}s +\frac{b_1+p(p(b_2-2p(p+9))+40)}{(p+4)(p+5)} \right) \nonumber
\end{eqnarray}
and they correspond to the terms ${\cal R}^4$ and $\partial^4 {\cal R}^4$ in the tree level effective action. The coefficients $b_1,b_2$ are not fixed by the analysis of \cite{Alday:2018pdi} and we will not fix them here. In this paper we will {\it assume} the vertices at tree level are given, and we will give a systematic way to construct the genus one amplitude from them. 
  
\section{Structure at one loop}
\label{loopstructure}
\subsection{Unitarity method on $AdS$}
The basic idea of the $AdS$ unitarity method developed in \cite{Aharony:2016dwx} is the following. Knowing the correlator at order $1/c$ we can compute the corresponding anomalous dimension for the double-trace operators $\gamma^{\text{tree}}_{n,\ell}$. This together with the explicit structure of the conformal block decomposition allows to compute  the $\log^2U$ piece of the correlator to order $1/c^2$:
\begin{equation}
\left. {\cal H}^{\text{loop}}(U,V)\right|_{\log^2U} = \frac{1}{8}\sum_{n,\ell} a^{\text{MFT}}_{n,\ell} \left(\gamma^{\text{tree}}_{n,\ell}\right)^2 g_{n,\ell}(U,V).
\end{equation}
where $g_{n,\ell}(U,V)$ is a short-hand notation for the conformal blocks of double-trace operators at zeroth order. Via crossing this contribution fixes the part of the answer proportional to $\log^2V$. Then \cite{Alday:2016njk} allows to reconstruct the full CFT-data from this piece, and in principle the whole correlator, up to certain ambiguities with finite support in the spin. In the language of \cite{Caron-Huot:2017vep} the piece proportional to $\log^2V$ gives the whole double-discontinuity to order $1/c^2$, from where the full CFT-data can be reconstructed through an elegant inversion formula. 

The case of ${\cal N}=4$ SYM at strong coupling is more complicated.  Due to its $R-$symmetry, there are several double-trace operators with the same twist and spin at large central charge:
\begin{equation}
[{\cal O}_2,{\cal O}_2]_{n,\ell},[{\cal O}_3,{\cal O}_3]_{n-1,\ell},\cdots , [{\cal O}_{n+2},{\cal O}_{n+2}]_{0,\ell}.
\end{equation}
and the sum over conformal blocks should include a sum over species for each $(n,\ell)$. Hence the correct expression for the piece proportional to $\log^2U$ is
\begin{equation}
\left. {\cal H}^{\text{loop}}(U,V)\right|_{\log^2U} = \frac{1}{8}\sum_{n,\ell} a^{\text{MFT}}_{n,\ell} \langle \left(\gamma^{\text{tree}}_{n,\ell}\right)^2\rangle  g_{n,\ell}(U,V).
\end{equation}
 where the square of the anomalous dimension has been replaced by its weighted average and the index $p=2$ has been suppressed in the OPE coefficients. In order to compute the weighted average we need to solve a mixing problem. For a given twist $4+2n$, this can be done by considering the family of correlators $\langle {\cal O}_2{\cal O}_2{\cal O}_p{\cal O}_p \rangle$, with $p=2,3,\cdots,n+2$. It can then be shown
 \begin{equation}
\langle \left(\gamma^{\text{tree}}_{n,\ell}\right)^2\rangle = \sum_{p=2}^{n+2} \langle \gamma^{\text{tree}}_{n,\ell} \rangle_{p}\langle \gamma^{\text{tree}}_{n,\ell} \rangle_{p}
\end{equation}
where $\langle \gamma^{\text{tree}}_{n,\ell} \rangle_{p}$ is the averaged anomalous dimension that follows from the correlator $\langle {\cal O}_2{\cal O}_2{\cal O}_p{\cal O}_p \rangle$ to order $1/c$. Note that this averaged anomalous dimension is the only information we can compute from the correlator, in the presence of more than one species. This justifies the figure in the introduction.  

Let us now discuss in detail the structure of ${\cal M}^{\text{loop}}(s,t)$. We will star by discussing ${\cal M}^{\text{loop-sugra}}(s,t)$, the result without stringy corrections. Then we will show how to systematically construct the stringy corrections from the result at tree level ${\cal M}^{\text{tree}}_p(s,t)$. 

\subsection{Loop supergravity}

In the following we give the full answer of the loop supergravity result in Mellin space. We start by recalling the $\log^2V$ coefficient of the answer in space time. This can be extracted from \cite{Aprile:2017bgs,Alday:2017vkk} and takes the form
\begin{eqnarray}
\label{g2dd}
 {\cal G}^{(2)}(U,V)\Big|_{\log^2V} =\frac{1}{ z\zb(\zb-z)} D(D-2)\bar D (\bar D-2)\hat {\cal G}^{(2)}(z,\zb) \label{dDisc2appendix}
\end{eqnarray}
where $D=z^2 \partial_z(1-z)\partial_z$ and 
\begin{eqnarray}
 \hat {\cal G}^{(2)'}(z,\zb) &=& R_0(z,\zb) + R_1(z,\zb)\left( \log z - \log \zb \right)+R_2(z,\zb)\left( \log z + \log \zb \right) \\
 & & + R_3(z,\zb) \left(\Li_2(1-z)-\Li_2(1-\zb) \right)+ R_4(z,\zb) \left(\Li_2(1-\frac{1}{z})-\Li_2(1-\frac{1}{\zb}) \right) \nonumber \\
 & & + \frac{1-\zb}{8U} \Li_2(1-z) - \frac{1-z}{8U} \Li_2(1-\zb) \nonumber
\end{eqnarray}
The rational functions $R_i(z,\bar z)$ where given in \cite{Alday:2017vkk},and can be found in appendix \ref{ddapp}. In \cite{Aprile:2017bgs} it was shown how to complete this piece to the full answer in space-time, assuming the answer has a specific structure in terms of transcendental functions. This result was confirmed in \cite{Alday:2017vkk}. In the following we will show that the answer has a remarkably simple structure in Mellin space. We start by making the following observation. ${\cal G}^{(2)}(U,V)$ contains a piece that behaves like  $\log^2V \log^2 z \sim \log^2V \log^2 U$. This can only arise from simultaneous poles in $s$ and $t$ in Mellin space. Crossing symmetry then implies the following symmetric structure
\begin{equation}
{\cal M}^{loop-sugra}(s,t) = \sum_{m,n=2} \left( \frac{c_{m n}}{(s-2m)(t-2n)} +\frac{c_{m n}}{(t-2m)(u-2n)} +\frac{c_{m n}}{(u-2m)(s-2n)} \right) + \cdots 
\end{equation}
with $c_{mn}=c_{nm}$ and recall $s+t+u=4$. The dots represent single poles and regular terms, to be addressed below. We then proceed as follows. Given $c_{mn}$ one can perform the corresponding residue integrals and compute the piece proportional to $\log^2U \log^2 V$ of the corresponding space-time answer. We obtain
\begin{equation}
{\cal G}^{(2)}(U,V) = \sum_{m,n=2} \frac{U^m V^{n-2} \Gamma (m+n)^2 c_{mn}}{16 \Gamma (m-1)^2 \Gamma (n-1)^2} \log^2U \log^2V + \cdots
\end{equation}
This should be matched to the known result, which can be read off from (\ref{g2dd}). It is convenient to expand it in powers of $U$:
\begin{equation}
\left. {\cal G}^{(2)}(U,V)\right|_{\log^2U \log^2V} = \frac{6\left(V^2+4 V+1\right)}{(1-V)^6}U^2 +\frac{12 \left(50 V^3+313 V^2+178 V+5\right)}{(1-V)^8}U^3  + \cdots
\end{equation}
This allows to find $c_{2n},c_{3n},\cdots$. After some work we were able to guess a closed form expression for $c_{mn}$. The coefficients are indeed symmetric under $m \leftrightarrow n$ and are given by
\begin{equation}
\label{cmn}
c_{mn} = \frac{p^{(6)}(m,n)}{5(m+n-5)_5}
\end{equation}
where $p^{(6)}(m,n)$ is a polynomial of degree six, given by

\begin{eqnarray}
p^{(6)}(m,n) &=&  30 m^2 n^2 (m+n)^2-10 m n \left(7 m^3+36 m^2 n+36 m n^2+7 n^3\right)-296  (m+n)+64 \nonumber \\
& &+ \left(44 m^4+548 m^3 n+1152 m^2 n^2+548 m n^3+44 n^4\right)\\
& & -2 \left(128 m^3+631 m^2 n+631 m n^2+128 n^3\right)+12  \left(37 m^2+90 m n+37 n^2\right) \nonumber
\end{eqnarray}
One can explicitly check that this polynomial vanishes for $m=n=2$; $m=2,n=3$ and $m=3,n=2$, so that $c_{mn}$ is finite in these cases. 

Having fixed the simultaneous poles of the Mellin amplitude, we can now turn to single poles in $s$ and $t$. A single pole in $s$ generates a term proportional to $\log^2U$, and hence it should be captured by (\ref{g2dd}), upon crossing symmetry. We find the following remarkable result: the whole double discontinuity (\ref{g2dd}) is reproduced by the simultaneous poles and hence single poles are absent. In appendix \ref{ambiguities} we furthermore show that regular terms are also absent, except for a single constant ambiguity. Hence, the full answer is given by
\begin{equation}
{\cal M}^{loop-sugra}(s,t) = \sum_{m,n=2} \left( \frac{c_{m n}}{(s-2m)(t-2n)} +\frac{c_{m n}}{(t-2m)(u-2n)} +\frac{c_{m n}}{(u-2m)(s-2n)} \right)
\end{equation}
with $c_{mn}$ given above.

\subsubsection{Flat space limit}
As mentioned in the introduction, the Mellin amplitude ${\cal M}(s,t)$ has the interpretation of a scattering amplitude on $AdS$. A strong indication that this must be the case is that in the limit in which the radius of $AdS$ becomes very large, and the Mellin variables are rescaled accordingly, one recovers the flat space scattering amplitude \cite{Penedones:2010ue}. For the case at hand \cite{Goncalves:2014ffa}
\begin{equation}
\lim_{R \to \infty} {\cal M}(R^2 s, R^2 t) \sim \int_0^{\infty} \beta^5 e^{-\beta} A(2\beta s,2\beta t)
\end{equation}
where $R$ is the radius of $AdS$ and $A(s,t)$ is the 10D flat scattering amplitude. See \cite{Alday:2017xua} for a derivation of the precise relation, following \cite{Chester:2018dga}, taking into account the graviton polarisations. In the following we will consider the large $s,t$ limit of ${\cal M}^{loop-sugra}(s,t)$. This is not straightforward, since its definition involves a double sum. Let us consider one of the three contributions
\begin{equation}
M_{part}(s,t)= \sum_{m,n=2} \frac{c_{m n}}{(s-2m)(t-2n)} 
\end{equation}
The first observation is that this sum is divergent. In order to obtain a convergent sum we take derivatives w.r.t. $s$ and $t$. 
\begin{equation}
\partial_{s} \partial_t M_{part}(s,t)= \sum_{m,n=2} \frac{c_{m n}}{(s-2m)^2(t-2n)^2} 
\end{equation}
the second observation is that any finite number of poles decays like $1/(s^2 t^2)$ for large $s,t$. However, from the explicit result for the flat space supergravity amplitude we expect a softer decay. Hence the enhanced behaviour arises from the tail of the sums: namely the regions where $n$ and $m$ are large and of the same order of $s,t$. Furthermore, if we are only interested in the leading  large $s,t$ behaviour we can replace the sums by integrals. We are led to
\begin{equation}
\partial_{s} \partial_t M_{part}(s,t) \approx \int_0^\infty dm dn \frac{6 m^2 n^2}{(m+n)^3 (s-2 m)^2 (t-2 n)^2} 
\end{equation}
the integrals can be performed and the answer written in terms of logarithms. Since we would like to compare our answer to the 10D box integral below, we will work in a region where logarithms are real in the Euclidean region, $s,t<0$. We obtain the following result
\begin{eqnarray}
\partial_{s} \partial_t M_{part}(s,t) &=& \frac{3 s t \left(s^2-4 s t+t^2\right)}{4 (s+t)^5}\log^2\frac{-s}{-t}-\frac{9 s t \left(s^2-t^2\right)}{2 (s+t)^5}\log\frac{-s}{-t} \\
& &+ \frac{3 \left(-s^4+2 \left(4+\pi ^2\right) s^3 t+2 \left(9-4 \pi ^2\right) s^2 t^2+2 \left(4+\pi ^2\right) s t^3-t^4\right)}{8 (s+t)^5} \nonumber
\end{eqnarray}
This precisely agrees, up to an overall factor entering the proper flat space limit, with the double derivative of the box function in ten dimensions $I(s,t) $
\begin{equation}
\partial_{s} \partial_t M_{part}(s,t)  = 45 \partial_{s} \partial_t I(s,t) 
\end{equation}
where
\begin{equation}
I(s,t) = \frac{1}{120} \Big( \frac{s^2t^2}{u^3} \left(\log^2\frac{-s}{-t}+\pi^2\right)-(s-t)\left(\frac{st}{u^2}+\frac12\right)\log\frac{-s}{-t}
+ u \log \frac{\sqrt{-s}\sqrt{-t}}{\Lambda^2}-\frac{st}{u}+a \Lambda^2+b u\Big)
\end{equation}
and here $s+t+u=0$. This is a non-trivial check of the Mellin expression for the loop supergravity result. Note that this match, together with crossing symmetry, strongly constraints the possible regular terms that we can add to the Mellin amplitude in the loop supergravity approximation. The only possibility is a constant. In appendix \ref{ambiguities} we perform an independent check of this fact. 
 
\subsection{Stringy corrections} 

In this section we give an algorithm to determine the stringy corrections to the Mellin amplitude at loop order for $p=2$, assuming the Mellin amplitude, for all $p$, to tree level. The recipe consists on three steps. 

\subsubsection*{Step 1: determining the average anomalous dimensions to order $1/c$}
The first step to solve the mixing problem is to determine the average anomalous dimensions from a given vertex in the correlator $\langle 22 pp \rangle$. The supergravity solution leads to the following contribution
\begin{equation}
\langle \gamma^{\text{tree-sugra}}_{n,\ell} \rangle_{p} = -\frac{(n+1)_4(n+\ell+2)_4(p-1) p^2 (p+1)  \Gamma (n+p+3) \Gamma (n-p+\ell+4)}{12 (\ell+1)^2 (2 n+\ell+6)^2 \Gamma (n-p+3) \Gamma (n+p+\ell+4)}
\end{equation}
Note that although the anomalous dimensions of individuals eigenstates must be independent of $p$, the weighted average is certainly not. For the vertices we proceed as follows.  Consider a particular vertex $V_p(s,t)$, which is a polynomial in $s,t$ and symmetric under $t \leftrightarrow u$. First we write it as a linear combination of the Mellin functions $M_{p,L}^{(q)}(s,t)$ presented in appendix \ref{basis}, where the functions in the $p-$frame must be used
\begin{equation}
V_p(s,t)  = c_0^{(q)}M_{p,0}^{(q)}(s,t) +c_2^{(q)} M_{p,2}^{(q)}(s,t)+ \cdots + c_{L_{max}}^{(q)} M_{p,L_{max}}^{(q)}(s,t).
\end{equation}
Here $L_{max}$ is given by the degree in $t$ of the polynomial $V_p(s,t)$. Once the coefficients $c_L^{(q)}$ are determined, it is straightforward to compute the anomalous dimension arising from this vertex. This is given by
\begin{equation}
\langle \gamma^{V}_{n,\ell}\rangle_p  = 2 \frac{p}{\Gamma(p-1)} \frac{c^{MFT}_{2,n,\ell}}{c^{MFT}_{p,n,\ell}} \sum_{q,L} c_L^{(q)} \rho^{L,q}_{n,\ell}
\end{equation}
where $c^{MFT}_{p,n,\ell}$ are the MFT OPE coefficients given in (\ref{MFTOPE}) and $ \rho^{L,q}_{n,\ell}$ is given in appendix \ref{basis}. 
Let us work out a simple example. Take $V_p(s,t) = \kappa_1(p)$, just a function of $p$, with no dependence on the Mellin variables. First we write $1$ as a linear combination of the functions $M_{p,L}^{(q)}(s,t)$ in the $p-$frame
\begin{equation}
 1 = -\frac{1}{4}(-1)^p M_{p,0}^{(p-2)}(s,t)
\end{equation}
Hence $V_p(s,t) = \kappa_1(p)$ leads to the following anomalous dimension
\begin{eqnarray}
\langle \gamma^{V}_{n,\ell}\rangle_p &=& - \kappa_1(p)\frac{(-1)^p p}{2\Gamma(p-1)} \frac{c^{MFT}_{2,n,0}}{c^{MFT}_{p,n,0}} \rho^{0,p-2}_{n,0} \\
&=&-\frac{\kappa_1(p) (-1)^p (n+2)^2 (n+3)^3 (n+4)^2 p}{(2 n+5) (2 n+7)(p+2)(p+3)}\sqrt{\frac{(n+1)^3  (n+5)^3 (p-1) ((n+3)^2-p^2)}{48  (p+1)}} \nonumber
\end{eqnarray}
with the results in appendix \ref{basis} it is in principle straightforward to compute the anomalous dimension due to any polynomial vertex. As proven in appendix \ref{inversion}, it is also possible to give an integral formula for the anomalous dimension given a specific vertex $V(s,t)$. Assume for simplicity $V_p(s)$ is solely a function of $s$. Then only operators with spin zero will acquire an anomalous dimension given by 
\begin{equation}
\langle \gamma^{V}_{n,0}\rangle_p =  -\frac{c^{MFT}_{2,n,\ell}}{c^{MFT}_{p,n,\ell}} \frac{(-1)^p(n+3)}{4(2n+5)(2n+7)} \int_{-i \infty}^{i \infty} ds \frac{\Gamma \left(\frac{s}{2}-n-2\right) \Gamma \left(\frac{s}{2}+n+4\right)}{\Gamma \left(\frac{s}{2}-1\right)^2} V_p(s)
\end{equation}

\subsubsection*{Step 2: Taking the square}
Once we have computed the averaged anomalous dimensions to order $1/c$ we can consider specific contributions proportional to a quartic vertex times supergravity or two quartic  vertices. These contributions were denoted by $\langle \gamma^2 \rangle^{sugra | {\cal R}^4}_{n,\ell}$, $\langle \gamma^2 \rangle^{{\cal R}^4 | {\cal R}^4}_{n,\ell}$, etc in \cite{Alday:2018pdi} and are represented by triangle and bubble loop diagrams, see figure.
\begin{figure}[h]
\includegraphics[scale=0.32]{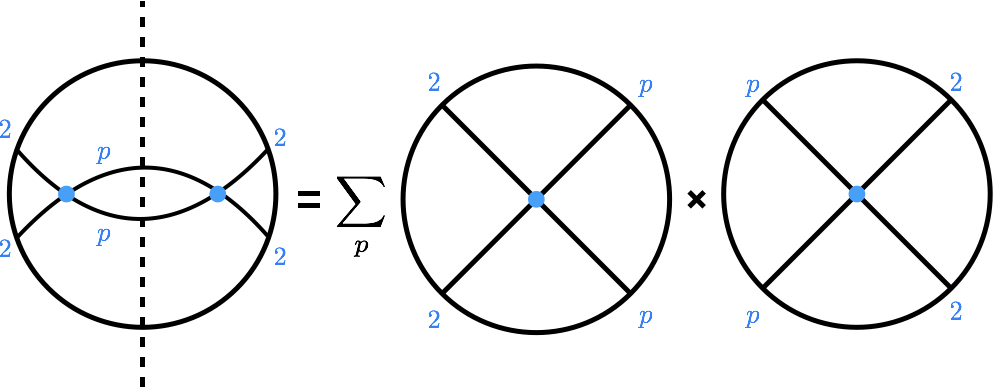}
\centering
\label{fig:bubble22pp}
\end{figure}
Following that notation, given two vertices $V,V'$, we have\footnote{For different vertices there is two equivalent contributions, so we need to multiply by a factor of $2$.}
\begin{equation}
\langle \gamma^2 \rangle^{V | V'}_{n,\ell} =\sum_{p=0}^{n+2} \langle \gamma^{V}_{n,\ell}\rangle_p \langle \gamma^{V'}_{n,\ell} \rangle_p
\end{equation}
The contribution $\langle \gamma^2 \rangle^{sugra | sugra}_{n,\ell}$ leads to the loop supergravity result, considered previously. Here we will consider the rest of the contributions. Let us focus in the example above, $V_p(s,t) = \kappa_1(p)$  and compute the contributions $\langle \gamma^2 \rangle^{sugra | V},\langle \gamma^2 \rangle^{V | V}$. We obtain
\begin{eqnarray}
\langle \gamma^2 \rangle^{sugra | V}_{n,\ell} &=&\sum_{p=0}^{n+2} \kappa_1(p) \frac{(n+1)_5^2(n+2)_3(p-1) p^2 (n-p+3)}{24 (2 n+5) (2 n+7) (p+2) (p+3)}\\
\langle \gamma^2 \rangle^{V | V}_{n,\ell} &=& \sum_{p=0}^{n+2} \kappa_1(p)^2 \frac{(n+1)_5^3 (n+2)_3(n+3)^2 (p-1) p^2 (n-p+3) (n+p+3)}{48 (2 n+5)^2 (2 n+7)^2 (p+1) (p+2)^2 (p+3)^2}
\end{eqnarray}
In order to perform these sums we need to specify $\kappa_1(p)$. In all the examples studied so far this always contains the factor $(p+1)_3$. For instance, to order $1/\lambda^{3/2}$, the leading stringy correction, the vertex is exactly of this form, and $\kappa_1(p)=\zeta_3 (p+1)_3$ is fixed by matching with the Virasoro-Shapiro amplitude in the flat space limit. One can perform the above sums in more generality, but let's take $\kappa_1(p)=(p+1)_3$  for definiteness. In this case we obtain
\begin{eqnarray}
\langle \gamma^2 \rangle^{sugra | V}_{n,\ell} &=&\frac{(n+1)^3 (n+2)^4 (n+3)^5 (n+4)^4 (n+5)^3}{720 (2 n+5) (2 n+7)}\\
\langle \gamma^2 \rangle^{V | V}_{n,\ell} &=&\frac{(n+1)^4 (n+2)^5 (n+3)^7 (n+4)^5 (n+5)^4}{3360 (2 n+5) (2 n+7)}
\end{eqnarray}
Note that the results are simple rational functions, with only single poles. Requiring this to be true puts some constraint on the possible $p$-dependence of the vertices at tree level. We will not prove that this is indeed the case, but we note that for the leading stringy corrections this is implied by the flat space limit. Furthermore, this is also strongly suggested  by the low energy expansion of the genus one string amplitude in flat space \cite{Green:2008uj}.

\subsubsection*{Step 3: Reconstructing the loop Mellin amplitude}
The final step is to reconstruct the contribution to the loop amplitude from a given $\langle \gamma^2 \rangle^{sugra | V}_{n,\ell},\langle \gamma^2 \rangle^{V | V'}_{n,\ell}$. For this, we perform the inverse of the procedure in step one. First, we write $\langle \gamma^2 \rangle^{V | V'}_{n,\ell}$ as a linear combination of the insertions $\rho_{n,\ell}^{L,q}$  given in appendix \ref{basis}.
\begin{equation}
\label{gamma2dec}
\langle \gamma^2 \rangle^{V | V'}_{n,\ell} = \sum_{q,L} c^{(q)}_{L} \rho_{n,\ell}^{L,q}
\end{equation}
The upper limit in $L$ is given by the minimum spin of the two vertices $V,V'$. The upper limit in $q$ is set by the behaviour of $\langle \gamma^2 \rangle^{V | V'}_{n,\ell}$ at large $n$. Having found the decomposition (\ref{gamma2dec}) we can readily write down the polar part of the Mellin amplitude at loop level:
\begin{equation}
{\cal M}^{(loop)}_{V | V'}(s,t) = \frac{1}{4} \sum_{q,L} c^{(q)}_{L}  M_L^{(q)}(s,t) \psi_0\left(2-\frac{s}{2} \right) + \text{crossed}
\end{equation}
where 'crossed' denotes the other two contributions such that the full amplitude is symmetric in $(s,t,u)$. This answer is crossing symmetric and reproduces the correct double-discontinuity. For instance, for the two diagrams considered above we obtain
\begin{eqnarray}
{\cal M}^{(loop)}_{sugra | V}(s,t) &=& \frac{1}{14} \left( -126 s^4+1288 s^3-5544 s^2+11552 s-9600 \right) \psi_0\left(2-\frac{s}{2} \right) + \text{crossed} \nonumber \\
{\cal M}^{(loop)}_{V | V}(s,t) &=& \frac{3^3 5}{8}\left( -462 s^7+11627 s^6-134274 s^5+908180 s^4-3841208 s^3+10071488 s^2 \right. \nonumber \\
& & \left. -15053056 s+9838080\right) \psi_0\left(2-\frac{s}{2} \right)+ \text{crossed} 
\end{eqnarray}
Note that to this answer we can still add regular, polynomial terms, completely symmetric in all Mellin variables. As we discuss in the following to a given order in $1/\lambda$ the degree of the possible ambiguities can be actually determined. 

 \subsection{Regularisation, UV divergences and ambiguities}
 
Some of the sums involved in defining the Mellin amplitude at loop order/genus-one are actually divergent. It is instructive to look at them in some detail. Consider for instance:
\begin{equation}
M_{div}(t)= \sum_{n=2} \frac{1}{t-2n}.
\end{equation}
This sum can be regulated by using zeta-function regularisation. An equivalent way is by taking derivatives w.r.t. $t$ and then performing the sum
  \begin{equation}
M_{div}'(t)= -\sum_{n=2} \frac{1}{(t-2n)^2} =- \frac{1}{4} \psi ^{(1)}\left(2-\frac{t}{2}\right)
 \end{equation}
 which leads to 
  \begin{equation}
M_{div}(t)= \frac{1}{2} \psi ^{(0)}\left(2-\frac{t}{2}\right) + C_0
 \end{equation}
 where $C_0$ is a constant of integration. Similar sums can be regulated in exactly the same way, but we may need to take more than one derivative. For instance
\begin{equation}
\sum_{n=2} \frac{n}{t-2n} = \frac{1}{4} t \psi ^{(0)}\left(2-\frac{t}{2}\right) + C_0 + C_1 t
\end{equation}
The presence of divergent sums in the one-loop Mellin amplitude is a manifestation of the UV divergences expected in the low energy expansion under consideration (loop supergravity plus a tower of stringy corrections). As seen for the two sums above, they can be regularised, but at the expense of introducing an ambiguity $C_0 + C_1 t$. Upon symmetrisation and recalling $s+t+u=4$, this leads to an ambiguity which is a constant.  This has exactly the same form as the correction due to a quartic vertex ${\cal R}^4$.  

A natural question to ask is, for a given divergent sum, what are the possible ambiguities. In the computation above we could have taken more derivatives, which would have introduced more constants of integration, and lead to a higher order polynomial in the Mellin variables.  The inversion formula of \cite{Caron-Huot:2017vep} explains why this is not possible and sets a limit in the order of the polynomial. Note that a given sum over poles produces a specific double-discontinuity in space time. Given this double-discontinuity one can compute the CFT-data for double-trace operators, for instance their anomalous dimension, to this order. Note that the double-discontinuity is blind to the polynomial ambiguities, but the anomalous dimension of low-spin operators is not. The fact that the inversion formula should produce the correct result in the region where it converges sets a bound on the polynomial ambiguities we can add.\footnote{The same conclusions can be drawn from large spin perturbation theory, if one assumes the CFT data can be pushed down to the radius of convergence in $1/\ell$, which is the case in this holographic context.}. As a result, for the sums above the ambiguous terms have to be the ones considered. 
   
Before concluding, let us mention a toy-model to understand how stringy corrections cure the UV divergences present in loop supergravity, at the level of the Mellin amplitude. Very roughly, string theory introduces a soft cut-off in the momenta, adding a factor $e^{-\alpha' p^2}$ in internal propagators for a given scattering process. This would suggest the following prescription:
\begin{equation}
M_{sugra}(t)= \sum_{n=2} \frac{n}{t-2n} \to M_{string}(t)= \sum_{n=2} \frac{n e^{-\alpha' n} }{t-2n}
\end{equation}
this toy model has the correct large $n$ behaviour at each order in $\alpha'$. Note that the sum $M_{string}(t)$ is now perfectly convergent for finite $\alpha'$. Furthermore, its corresponding double discontinuity produces a CFT data convergent for all values of the spin. However, if one performs the sum and then expands in $\alpha'$ each term, when plugged into the inversion formula, produces divergences for higher and higher values of the spin. More precisely, at order $\alpha'^q$ the inversion formula/CFT data converges only for $spin > q+1$. 

Let us now imagine we first expand in $\alpha'$ and then perform the sums, regularising them as above. To each order in $\alpha'$ we would obtain
\begin{eqnarray}
 \sum_{n=2} \frac{n}{t-2n} &=&  \frac{1}{4} t \psi ^{(0)}\left(2-\frac{t}{2}\right) + A_0 + A_1 t \\
 \sum_{n=2} \frac{n^2}{t-2n} &=&  \frac{1}{8} t^2 \psi ^{(0)}\left(2-\frac{t}{2}\right) + B_0 + B_1 t +B_2 t^2
\end{eqnarray}
and so on. But for the toy model at hand this can be precisely compared with the small $\alpha'$ expansion of the exact answer. We obtain
\begin{eqnarray}
\sum_{n=2} \frac{n e^{-\alpha' n} }{t-2n}= -\frac{1}{2 \alpha'} +  \frac{1}{4} t \psi ^{(0)}\left(2-\frac{t}{2}\right) + \frac{1}{4} (t \log \alpha'+\gamma_E  s+3) + \cdots
\end{eqnarray}
A very interesting feature is the appearance of the divergent term $-\frac{1}{2 \alpha'}$. This has exactly the form  of the known counterterm to regularise the loop supergravity computation. Generically the structure of these enhanced terms is as follows
\begin{eqnarray}
\sum_{n=2} \frac{(2n)^q e^{-\alpha' n} }{t-2n} \sim \frac{1}{\alpha'^q} + \cdots + \frac{t^{q-1}}{\alpha'}+ t^q \psi ^{(0)}\left(2-\frac{t}{2}\right)
\end{eqnarray}
Note also the appearance  of regular terms proportional to $\log \alpha'$. Still, we would like to stress that this should only be seen as a toy model. For instance, a bug of this model is that it produces a non-trivial Mellin amplitude to order $\alpha'$, which is not present in the full solution. 

\subsection{Summary: full structure at one-loop}
The low energy expansion of the one-loop string amplitude on $AdS_5 \times S^5$ takes the following form
\begin{equation}
{\cal M}^{loop}(s,t,u) = \hat {\cal M}_{polar}(s,t,u) + \hat {\cal M}_{polar}(t,u,s) + \hat {\cal M}_{polar}(u,s,t) + R(s,t,u)
\end{equation}
with
\begin{equation}
\hat {\cal M}_{polar}(s,t,u) = \sum_{m,n=2} \frac{c_{mn}}{(s-2m)(t-2n)}  + \sum_{n=3,5,6,\cdots} \frac{1}{\lambda^{n/2}} P^{(n+1)}(s,t,u) \psi_0\left(2-\frac{s}{2} \right)
\end{equation}
with $c_{mn}$ given in (\ref{cmn}) and where $P^{(n+1)}(s,t,u)$ are polynomials in the Mellin variables symmetric under $t \leftrightarrow u$ and of total degree $n+1$. These polynomials can be constructed following the prescription outlined above. At each order in $1/\lambda$ we have the freedom to add regular polynomial terms:
\begin{equation}
 R(s,t,u)= \alpha_0 + \frac{1}{\lambda^{3/2}} p^{(1)}(s,t,u)+ \frac{1}{\lambda^{5/2}}  p^{(2)}(s,t,u) + \cdots
\end{equation}
fully symmetric in the Mellin variables and with bounded degree, as discussed above. These polynomials cannot be fixed from crossing symmetry alone. Note that based in the toy-model above we expect these regular terms to contain also pieces proportional to $\log \alpha'$. It is illuminating to compare this answer to the corresponding expansion in flat space, given by \cite{Green:2008uj,DHoker:2015gmr}
\begin{equation}
A^{genus-1}(s,t,u) = A^{genus-1}_{an}(s,t,u) + A^{genus-1}_{non-an}(s,t,u) 
\end{equation}
where the analytic contribution is given by
\begin{eqnarray}
A^{genus-1}_{an}(s,t,u) =  \frac{\pi}{3} \left(1+ 0 \sigma_2+ \alpha'^3 \frac{\zeta_3}{192} \sigma_3 +\cdots \right) 
\end{eqnarray}
while the non-analytic part is given by 
\begin{eqnarray}
A^{genus-1}_{non-an}(s,t,u) &=&  A^{loop-sugra}(s,t,u)  + \left( - \left(\frac{\alpha'}{4}\right)^4  \frac{4\zeta_3 \pi}{45} s^4 \log \left(-\frac{\alpha' s}{\mu_4} \right) \right. \\
& & \left. - \left(\frac{\alpha'}{4}\right)^6  \frac{\zeta_5 \pi}{2520} (87 s^6+s^4(t-u)^2) \log \left(-\frac{\alpha' s}{\mu_6} \right) +  \text{crossed} + \cdots \right) \nonumber
\end{eqnarray}
We see that in the flat space limit the structure of the answer in $AdS$ reduces exactly to that in flat-space, where infinite sums over poles lead to branch cuts. Furthermore, note  that in the conventions of \cite{Green:2008uj} $A^{loop-sugra}(s,t,u)$ has an overall coefficient $\alpha'$.\footnote{The apparent mismatch in the powers of $\alpha'$ is due to different conventions between the CFT and string computation. For instance, from the results in \cite{Green:2008uj} one can explicitly check that the relative factor between the tree level supergravity result and the first non-analytic genus-one contribution is $g_s^2 \alpha'^7$. Using the $AdS/CFT$ dictionary this precisely reduces to $\frac{1}{c} \frac{1}{\lambda^{3/2}}$, which is exactly what we obtain.} Hence the analytic contribution is enhanced by a power of $1/\alpha'$ with respect to the non-analytic piece and can be interpreted as regulating the UV divergences that arise when computing the genus one-amplitude, exactly as explained in the toy model above. Note furthermore that the regular terms in the $AdS$ solution will include the terms proportional to $\log \alpha'$ and $\log \mu$ in the flat-space limit.  
 
\section{Conclusions} 

In the present paper we have studied non-planar corrections to the stress-tensor four-point correlator in ${\cal  N}=4$ SYM, to order $1/c^2$ and in a large 't Hooft coupling expansion. We have studied such correlator in Mellin space, showing that it displays a remarkable simple structure. Our motivation is that this should represent the genus one graviton amplitude for type IIB strings on $AdS_5 \times S^5$. This cannot be accessed by direct methods.  

Ours is a modest result, but the amplitude in Mellin space displays very interesting features. In the loop-supergravity approximation the amplitude is given by an infinite sum over simultaneous poles, plus a constant. Stringy corrections are given by sums over single poles, whose residues follow from the result at tree level (for more general amplitudes, including also KK modes). In addition we can add regular polynomial terms, whose degree is fixed at each order in $1/\lambda$. The result in Mellin space allows a direct comparison to the amplitude in flat space, and its elucidating to contrast the two. Furthermore, there is a direct relation between the UV divergences present in the loop-supergravity result (which stringy corrections cure) and the degree of the polynomial ambiguities mentioned above. It would be very interesting to make contact with recent impressive developments regarding the Mellin/Polyakov bootstrap, see \cite{Gopakumar:2016wkt,Gopakumar:2016cpb,Gopakumar:2018xqi,Ghosh:2018bgd,Mazac:2018ycv,Mazac:2018qmi}, where the issue of contact terms is very important. In the problem at hand we are dealing with a full-fledge holographic CFT and string theory and the flat space limit put some constraints on what the regular terms can be. 

As a byproduct we have also derived in inversion formula to compute the anomalous dimension to double-trace operators given a polynomial Mellin amplitude. The idea behind its derivation is the same as the idea that relates large spin perturbation theory to the usual inversion formula, see \cite{Alday:2017vkk}. However, the formula we have derived only works order by order in $t$, so we feel it is far from optimal. It would be interesting to make contact with related attempts \cite{Cardona:2018nnk,Cardona:2018dov,Zhou:2018sfz}.

There are many additional open questions that would be interesting to address. First, note that we have {\it assumed} the tree-level answer for the family of correlators $\langle 22pp \rangle$ is given beyond the supergravity approximation. It would be very interesting to understand how to fix the stringy corrections in this more general case, perhaps along the lines of \cite{Chester:2018aca}. In \cite{Alday:2018pdi} we have conjectured certain principles that strongly constraint the form of such corrections, but still leave some freedom. 

There has been a lot of progress in understanding which kind of functions can appear in the low energy expansion of string amplitudes in flat space, see for instance \cite{DHoker:2015wxz,Vanhove:2018elu,Broedel:2018izr,Zerbini:2018hgs}. These are special modular functions  with a very rich mathematical structure, see \cite{DHoker:2016mwo,Basu:2016kli,Brown:2017qwo}. It would be fascinating to understand how much of this structure extends to curve space-time. 

Another interesting question is the extension to higher genus. From the CFT perspective, higher trace operators will also enter into the game. A step towards this aim would be first to understand genus one corrections to more general correlators, involving the KK-modes of the graviton. At tree level and in the supergravity approximation a very nice structure arises   \cite{Rastelli:2017udc,Caron-Huot:2018kta}. It would be interesting to see if any of this structure survives at loop order and/or when including stringy corrections. 

\section*{Acknowledgements} 

This project has received funding from the European Research Council (ERC) under the European Union's Horizon 2020 research and innovation programme (grant agreement No 787185). I would like to thank A. Bissi for collaboration at the initial stage of this project and A. Bissi , M. Green, E. Panzer and A. Sinha for useful discussions. I would like to thank the S.N. Bose National Centre for Basic Sciences, Kolkata, for hospitality during the final stages of this work.  
 
\appendix

\section{A basis for polynomial Mellin amplitudes}
\label{basis}
An important ingredient used in this paper is a family of sums over super-conformal blocks, truncated in the spin and labelled by $L$, the highest spin entering the sum, and $q=0,1,\cdots$. These sums take the form
\begin{equation}
\label{specialsums}
S_L^{(q)}(z,\bar z)  = \sum_{\ell}^L \sum_n a_{n,\ell}^{\text{MFT}} \rho^{(L,q)}_{n,\ell} g_{n,\ell}(z,\bar z)
\end{equation}
where $a_{n,\ell}^{\text{MFT}}$ are the (square) OPE coefficients of the double-trace operators, in the mean field theory approximation, given by (\ref{MFTOPE}) for $p=2$. $g_{n,\ell}(z,\bar z)$ is a short-hand notation for $g_{\Delta_n,\ell}(z,\bar z)$, with $\Delta_n = 4+2n+\ell$, where the super-conformal blocks are given by
\begin{equation}
g_{\Delta,\ell}(z,\bar z) = (z \bar z)^{\frac{\Delta-\ell}{2}} \frac{z^{\ell+1} F_{\frac{\Delta+\ell+4}{2}}(z)F_{\frac{\Delta-\ell+2}{2}}(\bar z)-\bar z^{\ell+1} F_{\frac{\Delta+\ell+4}{2}}(\bar z)F_{\frac{\Delta-\ell+2}{2}}(z)}{z-\bar z}
\end{equation}
with $F_{\beta}(z)=~_2F_1(\beta,\beta,2\beta,z)$ the standard hypergeometric function. For $L=0$
\begin{equation}
\rho^{(0,q)}_{n,0}=\frac{(n+1)_5(n+3) \Gamma (n+q+6)}{(2 n+5) (2 n+7) \Gamma (n-q+1)  \Gamma (q+6)}
\end{equation}
while for $L=2$:
\begin{eqnarray}
\rho^{(2,q)}_{n,2}&=&\frac{(n+1)_7 (n+3)_3 \Gamma (n+q+8)}{(2 n+5) (2 n+7) (2 n+9) (2 n+11) \Gamma (n-q+1) \Gamma (q+8)}\\
\rho^{(2,q)}_{n,0}&=&\frac{2 q+5}{3 (2 q+9)}\rho^{(2,q)}_{n-1,2}
\end{eqnarray}
The expressions for all $L$ take the form
\begin{eqnarray}
\rho^{(L,q)}_{n,L}&=&  \frac{\Gamma \left(n+\frac{5}{2}\right) \Gamma (L+n+4) \Gamma (L+n+6) \Gamma (L+n+q+6)}{2^{L+2} \Gamma (n+1) \Gamma (n+3) \Gamma \left(L+n+\frac{9}{2}\right) \Gamma (L+q+6) \Gamma (n-q+1)}\\
\rho^{(L,q)}_{n,L-2j} &=& \rho^{(L,q)}_{n-j,L} \frac{\Gamma \left(q+\frac{5}{2}+j \right) \Gamma \left(L+q+\frac{7}{2}-j \right)}{\Gamma \left(q+\frac{5}{2}\right) \Gamma \left(L+q+\frac{7}{2}\right)} \kappa^{(L)}_j
\end{eqnarray}
for $j=1,2,\cdots,L/2$, where
\begin{equation}
\kappa^{(L)}_j= \frac{4 \Gamma \left(j+\frac{1}{2}\right) \Gamma \left(j+\frac{5}{2}\right)\Gamma (L+2) \Gamma \left(-j+L+\frac{3}{2}\right) \Gamma \left(-j+L+\frac{7}{2}\right)}{3 \pi  \Gamma (j+1) \Gamma \left(L+\frac{3}{2}\right) \Gamma \left(L+\frac{7}{2}\right) \Gamma (-j+L+2)}
\end{equation}
A feature of these insertions, to be used in the body of the paper, is the large $n$ behaviour, given by 
\begin{equation}
\rho^{(L,q)}_{n,\ell} \sim n^{2q+9+2L}
\end{equation}
In \cite{Alday:2018pdi} we characterised related sums as having a particularly simple structure, involving only rational functions and $\log(1-z), \log(1-\bar z)$. In this paper we find a more precise characterisation: there is a basis of polynomial Mellin amplitudes, symmetric under $t \leftrightarrow u$ and denoted as $M_L^{(q)}(s,t)$, such that their $\log U$ piece in space-time exactly coincides with $S_L^{q}(U,V)$. $M_L^{(q)}(s,t)$ takes the following form:
\begin{eqnarray}
M_L^{(q)}(s,t) = -\frac{2^{2-\frac{L}{2}} \Gamma \left(\frac{s}{2}-1\right) \Gamma \left(\frac{L}{2}+q+\frac{7}{2}\right)}{\Gamma \left(L+q+\frac{7}{2}\right) \Gamma \left(-q+\frac{s}{2}-1\right)}P^{(q)}_L(s,t)
\end{eqnarray}
where $P^{(q)}_L(s,t)$ is a polynomial of degree $L$ in $s,t,q$ and possesses the symmetry 
$$P^{(q)}_L(s,t)=P^{(q)}_L(s,4-s-t)$$
For the first few cases we obtain
\begin{eqnarray}
P^{(q)}_0(s,t) &=& 1 \\
P^{(q)}_2(s,t) &=& \frac{31 s^2+162 s t-208 s+162 t^2-648 t+624}{6} \\
& & + \frac{7 s^2+36 s t-28 s+36 t^2-144 t+64}{6} q+ \frac{2}{3} q^2 (s-4) \nonumber
\end{eqnarray}
For a given $L$ the polynomials can be determined as follows. The four-dimensional super conformal blocks are eigenfunctions of a quadratic Casimir
\begin{equation}
{\cal C} g_{n,\ell}(z,\bar z) = \lambda_{n,\ell} g_{n,\ell}(z,\bar z),~~~~~\lambda_{n,\ell}=(n+\ell+3)(n+\ell+4) +(n+1)(n+4)
\end{equation}
This Casimir has a specific action on Mellin space, as a difference operator: 
\begin{equation}
{\cal C} =  \frac{1}{4} \left(s^2-2 s t+8 s-2 t^2+8 t-16\right) -\frac{1}{4} (s-4)^2 T_s^{-}(1+T_t^+) + \frac{1}{4} (t-4)^2 T_t^-+\frac{1}{4} (s+t)^2 T_t^+
\end{equation}
where the shift operators $T^{\pm}_{s,t}$ shift the corresponding Mellin variable by $\pm 2$. Acting with the Casimir operator on the sums (\ref{specialsums}) will have the effect of multiplying $\rho_{n,\ell}^{L,q} \to \lambda_{n,\ell} \rho_{n,\ell}^{L,q}$.  This can be written as a linear combination of the $\rho_{n,\ell}^{L',q'}$ themselves with $q' = q+1,q,\cdots$ and $L'=L,L-2,\cdots$. This implies a difference equation relating ${\cal C} M_L^{(q)}(s,t)$ to a linear combination of the $M_{L'}^{(q')}(s,t)$ themselves. For any fixed $L$ this can be worked out, and leads to a difference equation from which the polynomials can be fixed. 

The functions $M_L^{(q)}(s,t)$ represent a basis for polynomials in $s,t$ symmetric under $t \leftrightarrow 4-s-t$, where the degree in $t$ is given by $L$. For instance, for $L=0$ we obtain
\begin{eqnarray}
M_0^{(0)}(s,t) = -4,~~~M_0^{(1)}(s,t) = -4 (\frac{s}{2}-2),~~~M_0^{(q)}(s,t) = -4 (\frac{s}{2}-2) (\frac{s}{2}-3) \cdots (\frac{s}{2}-q-1)
\end{eqnarray}
which clearly form a basis for polynomials in $s$. The above Mellin expressions are given in the 'frame' of the $\langle 2222 \rangle$ correlator. For the application in the body of the paper it will be useful to translate such Mellin expressions to the frame of the correlator $\langle 22pp \rangle$. If a given function of cross-ratios has a Mellin representation $M(s,t)$ with respect to $\langle 2222 \rangle$, then its Mellin representation in the $p-$frame is simply
\begin{eqnarray}
\left. M(s,t) \right|_{p-frame} = \frac{\Gamma(2-s/2)}{\Gamma(p-s/2)} M(s,t+2-p)
\end{eqnarray}
we will denote the above Mellin expressions in the $p-$frame as $M_{p,L}^{(q)}(s,t)$. As usual, for $p=2$ we suppress this index. 

Let us add the following observation. In the body of the paper it is important to reproduce space-time answers with a specific piece proportional to $\log^2U$. From the discussion above, together with basic properties of residues it follows
\begin{equation}
S_L^{(q)}(z,\bar z)  \log^2 U \leftrightarrow 2 M_L^{(q)}(s,t)  \psi_0\left( 2-\frac{s}{2} \right) 
\end{equation}
where $\psi_0(x)$ is the digamma function. 
 
\section{Inverting polynomial Mellin amplitudes}
\label{inversion}
The previous appendix gives a systematic way to compute the anomalous dimension given a polynomial Mellin amplitude of a fixed degree in $t$. We would like to derive an integral formula for this. Let us start with the simplest case $M(s,t) = M_0(s)$, a Mellin amplitude which is just a function of $s$. Following the discussion in the previous appendix, only operator with spin zero will acquire an anomalous dimension. We look for an inversion formula of the form
\begin{equation}
\gamma_{n,0} = \int_{-i \infty}^{i \infty} ds K_0(s)  M_0(s)
\end{equation}
Acting with the Casimir operator on both sides of this equation we obtain
\begin{eqnarray*}
\lambda_{n,0} \gamma_{n,0} = \int_{-i \infty}^{i \infty} ds K_0(s)  {\cal C} M_0(s) &=& \int_{-i \infty}^{i \infty} ds K_0(s) \frac{1}{2}\left( s(4+s) M_0(s) -(s-4)^2 M_0(s-2)  \right) \\
&=& \int_{-i \infty}^{i \infty} ds  \frac{1}{2}\left(s(4+s) K_0(s)-(s-2)^2 K_0(s+2) \right) M_0(s)
\end{eqnarray*}
where in the second line we shifted the contour of integration in the second term. This should be true for generic $M_0(s)$ and leads to 
\begin{eqnarray*}
\lambda_{n,0} K_0(s)= \frac{1}{2}\left(s(4+s) K_0(s)-(s-2)^2 K_0(s+2) \right)
\end{eqnarray*}
which gives a recursion relation from which we can determine $K_0(s)$ up to a multiplicative $n-$dependent factor:
\begin{equation}
K_n^{(0)}(s) = h_n \frac{\Gamma \left(\frac{s}{2}-n-2\right) \Gamma \left(\frac{s}{2}+n+4\right)}{\Gamma \left(\frac{s}{2}-1\right)^2}
\end{equation}
The overall coefficient can be fixed as follows. Using the basis of solutions given in appendix (\ref{basis}) for $L=q=0$ we require
\begin{equation}
2 \rho^{(0,0)}_{n,0} = -4 \int_{-i \infty}^{i \infty} K_n^{(0)}(s) ds.
\end{equation}
We find 
\begin{equation}
h(n) = - \frac{n+3}{4(2n+5)(2n+7)},
\end{equation}
where we have used the following result
\begin{equation}
\int_{-i \infty}^{i \infty} ds \frac{\Gamma\left(\frac{s}{2}+n+q\right)\Gamma\left(\frac{s}{2}-n-2\right)}{\Gamma\left(\frac{s}{2}-1\right)^2} = \frac{2 \Gamma (n+q+2)^2}{\Gamma (n+1)^2 \Gamma (q+2)}
\end{equation}
where for integer $n$ the contour includes the poles $s=4,6,\cdots,2(n+2)$. 

This method can be generalised to any polynomial Mellin amplitude. Let's pick the following basis of symmetric functions in $t,u$:
\begin{equation}
M(s,t) = M_0(s) + (t^2+u^2)M_2(s)+ (t^4+u^4)M_4(s) + \cdots
\end{equation}
Under the action of the Casimir
\begin{equation}
{\cal C} M(s,t) = \hat M_0(s) + (t^2+u^2)\hat  M_2(s)+ (t^4+u^4) \hat M_4(s) + \cdots
\end{equation}
Let us introduce a matrix notation in this basis, ${\bf M}(s)  = (M_0(s) ,M_2(s),\cdots)^T$ and the same for $\hat {\bf M}$. Then we have
\begin{equation}
\hat {\bf M}(s) = {\bf C}_0(s) {\bf M}(s)  +  {\bf C}_-(s) {\bf M}(s-2)
\end{equation}
where ${\bf C}_0(s),{\bf C}_-(s)$ are upper triangular matrices, which are calculable for truncations of any size:

\begin{eqnarray*}
{\bf C}_0(s) &=& \begin{bmatrix} 
    \frac{1}{2} s (s+4) & -\frac{1}{2} (s-8) s (s+4)-32 & \dots \\
   0  & \frac{1}{2} (s (s+8)+20) & \dots \\
    \vdots & \vdots &\ddots 
    \end{bmatrix},\\
    {\bf C}_-(s) &=& \begin{bmatrix} 
    -\frac{1}{2} (s-4)^2 & \frac{1}{2} (s-6) (s-4)^2 & \dots \\
   0  & -\frac{1}{2} (s-4)^2 & \dots \\
    \vdots & \vdots &\ddots 
    \end{bmatrix}
\end{eqnarray*}

Now we would like to construct a family Kernels $K_{m,n}(s)$ such that
\begin{eqnarray}
\label{KernelsallL}
\gamma_{n,\ell} &=& \int_{-i \infty}^{i \infty} ds \left( K_n^{(\ell,\ell)}(s) M_\ell(s) +  K_n^{(\ell,\ell+2)}(s) M_{\ell+2}(s) + \cdots \right) 
\end{eqnarray}
In matrix notation, we can consider an upper triangular and a diagonal matrices 
\begin{equation}
({\bf K}(s))_{i,j}=K_n^{(2i,2j)}(s),~~~{\bf J} = diag(\lambda_{n,0},\lambda_{n,2},\lambda_{n,4},\cdots)
\end{equation}
Acting with the Casimir on both sides of (\ref{KernelsallL}) and shifting the contour we arrive to the following relation
\begin{equation}
{\bf J} \, {\bf K}(s) =  {\bf K}(s) {\bf C}_0(s) +  {\bf K}(s+2) {\bf C}_-(s+2)
\end{equation}
This leads to an infinite set of relations for the Kernels $K_n^{(\ell,\ell+2m)}(s)$, $m=0,1,2,\cdots$ which can be solved iteratively. The solution to these relations takes the following form
\begin{equation}
K_n^{(\ell,\ell+2m)}(s) = \frac{\Gamma \left(n+\frac{5}{2}\right) \Gamma (\ell+n+4)}{2^{\ell+2} \Gamma (n+3) \Gamma \left(\ell+n+\frac{9}{2}\right)} \frac{\Gamma \left(\frac{s}{2}-n-2\right) \Gamma \left(\frac{s}{2}+n+4+\ell \right)}{\Gamma \left(\frac{s}{2}-1\right)^2} P^{(2m)}_{n,\ell}(s)
\end{equation}
where $P^{(2m)}_{n,\ell}(s)$ are polynomials in $s$ of degree $2m$, which are uniquely fixed by the recurrence relations and $P^{(0)}_{n,\ell}(s)=1$. For instance the second order polynomials $P^{(2)}_{n,\ell}(s)$ satisfy
\begin{eqnarray*}
(2 n-s+4) (2 \ell +2 n+s+8)P^{(2)}_{n,\ell}(s+2) +(s (2 \ell+s+8)+2 (\ell+2) (\ell+5)-2 \lambda_{n,\ell})P^{(2)}_{n,\ell}(s) =\\
= \frac{1}{3} (\ell+1) (\ell+2) \left(-3 (s-4) \lambda_{n,\ell}+\ell (3 \ell+25) s-10 \ell (\ell+7)+6 s^2+24 (s-4)\right)
\end{eqnarray*}
assuming they are second order polynomials in $s$, this relation fixes them uniquely.

\section{Double-discontinuity of loop supergravity}
\label{ddapp}

In the body of the paper we gave the double-discontinuity of the correlator to order $1/c^2$, in the supergravity approximation, in terms of five rational functions. These functions are given by
\begin{eqnarray*}
R_0(z,\zb) &=& \frac{U V^3 (3 V-7 (U+1))}{16 (z-\zb)^5}+\frac{V^2 (-4 U-3 V+15)}{48 (z-\zb)^3}+\frac{V \left(\frac{7 U}{3}-V-3\right)}{16 U (z-\zb)}-\frac{z-\zb}{16 U}\\
R_1(z,\zb) &=& \frac{V^2 \left(U^2-U+V-1\right)}{8 (z-\zb)^4}+\frac{U V^3 \left(U^2-U V+5 U-V+1\right)}{8 (z-\zb)^6}+\frac{(1-V) V}{8 U (z-\zb)^2}\\
& & -\frac{(1-U)^2+5 V}{96 (z-\zb)^2}+\frac{V (V+1)-2 (1-U)^2}{64 U V}+\frac{13}{192} \\
R_2(z,\zb) &=& \frac{U \left(1-U^2\right) V^2}{8 (z-\zb)^5}+\frac{V}{8 U (z-\zb)}+\frac{V (U (1-U)-6 (-U+V+1))}{96 (z-\zb)^3}\\
& & +\frac{(2 U-V-2) (z-\zb)}{64 U V}+\frac{1-U+V}{96 (z-\zb)}\\
R_3(z,\zb) &=& \frac{U V^2 (U-V-1)}{8 (z-\zb)^6}+\frac{V (U-V-1)}{8 U (z-\zb)^2}+\frac{V^2}{4 (z-\zb)^4}\\
R_4(z,\zb) &=& \frac{U^3 V^2 (U+V-1)}{8 (z-\zb)^6}
\end{eqnarray*}
Where we have used a mixed notation for the cross-ratios, with $U=z \bar{z}$, $V=(1-z)(1-\bar z)$.

\section{Regular terms ambiguities for loop supergravity}
\label{ambiguities}

In the body of the paper we have seen that the precise double discontinuity in the loop supergravity approximation implies a Mellin amplitude which is the sum over simultaneous poles in two Mellin variables. In particular, single poles are absent. Furthermore, matching with the flat space limit suggest that regular terms are also absent, except for a single constant ambiguity (the simplest quartic vertex) independent of the Mellin variables. In this appendix we perform an independent check of this fact. In order to do this we take as a starting point the Mellin expression, in terms of simultaneous poles, and compute the corresponding space-time answer. We will work to order $U^2$ and keep only the pieces proportional to $\log^2U$ and $\log U$. Note that the three terms will contribute to this. We would like to study the Mellin amplitude around $s=4$, keeping also regular terms. Let us study the three terms separately: 
\begin{eqnarray*}
\sum_{m,n=2} \frac{c_{mn}}{(s-2m)(t-2n)} &=&\frac{1}{s-4} \sum_n \frac{c_{2n}}{t-2n} + \sum_{m \neq 2,n} \frac{c_{mn}}{(4-2m)(t-2n)}\\
\sum_{m,n=2} \frac{c_{mn}}{(s-2m)(u-2n)} &=& -\frac{1}{s-4} \sum_n \frac{c_{2n}}{t+2n}  + \sum_n \frac{c_{2n}}{(t+2n)^2} - \sum_{m \neq 2,n} \frac{c_{mn}}{(4-2m)(t+2n)} \nonumber \\
\sum_{m,n=2} \frac{c_{mn}}{(t-2m)(u-2n)} &=& -\sum_{m,n=2} \frac{c_{mn}}{(t-2m)(t+2n)} 
\end{eqnarray*}
Up to terms that vanish as $s \to 4$. Some of sums involved are actually divergent. The presence of divergences is actually expected, and at the end of this appendix we explain how to regularise such sums. The final result can be written in terms of polygamma functions, although is quite cumbersome. Performing the Mellin integrals we arrive at the following expression 
 \begin{equation}
 {\cal G}^{(2)}(U,V) = U^2\left( \log^2 U h_0(V) + \log U h_1(V) + h_2(V) \right) + \cdots
 \end{equation}
 $$h_0(V) = \frac{6 \log V \left(-3 V^2+\left(V^2+4 V+1\right) \log V+3\right)}{(1-V)^6}$$
 and
 \begin{eqnarray*}
h_1(V) &=& \frac{6 \left(3 V^4-99 V^3-333 V^2-2 (V-1)^2 \left(V^2+4 V+1\right) \log (1-V)-99 V+3\right)}{(1-V)^8} \log^2 V\\
& & + \frac{72 \left(V^2+4 V+1\right)}{(1-V)^6} \zeta_3 + \frac{4 \left(-3 V^2+2 \left(V^2+4 V+1\right) \log V+3\right)}{(1-V)^6} \pi^2 \\
& & + \frac{18 (V+1) \left(-V^2+72 V+4 (V-1)^2 \log (1-V)-1\right)}{(V-1)^7} \log V \\
& & + \frac{24 (V (V+4)+1)}{(V-1)^6}  \log V \text{Li}_2(V) -\frac{(72 (V+1))}{(1-V)^5} \text{Li}_2(V) \\
& & + \frac{9 \left(27 V^2+86 V+27\right)}{2 (1-V)^6} -\frac{72 \left(V^2+4 V+1\right)}{(1-V)^6}\text{Li}_3(V)
 \end{eqnarray*}
 $h_2(V)$ will not be necessary for our purposes. With this expression at hand, we can perform a CPW decomposition and compute the anomalous dimension of leading twist (four) double trace operators. We find
 \begin{equation}
 \gamma_{0,\ell}^{loop-sugra} = 24\frac{7 \ell^4+74 \ell^3-553 \ell^2-4904 \ell-3444}{(\ell-1)(\ell+1)^3(\ell+6)^3(\ell+8)},~~~~\ell \geq 2
 \end{equation}
But this agrees precisely with the known result, see \cite{Aprile:2017bgs}. This still leaves the ambiguity of adding a regular term that only contributes to the anomalous dimension of spin zero operators. This exactly matches our intuition from the flat space limit.

\bibliographystyle{utphys} 
\bibliography{genusone}

\providecommand{\href}[2]{#2}\begingroup\raggedright\begin{thebibliography}{10}

\bibitem{Goroff:1985th}
M.~H. Goroff and A.~Sagnotti, ``{The Ultraviolet Behavior of Einstein
  Gravity},''
\href{http://dx.doi.org/10.1016/0550-3213(86)90193-8}{{\em Nucl. Phys.}
  {\bfseries B266} (1986) 709--736}.

\bibitem{vandeVen:1991gw}
A.~E.~M. van~de Ven, ``{Two loop quantum gravity},''
\href{http://dx.doi.org/10.1016/0550-3213(92)90011-Y}{{\em Nucl. Phys.}
  {\bfseries B378} (1992) 309--366}.

\bibitem{Vanhove:2009zz}
P.~Vanhove, \href{http://dx.doi.org/10.1142/9789814374552_0476}{``{On the
  ultraviolet behaviour of N=8 supergravity amplitudes},''} in {\em {On recent
  developments in theoretical and experimental general relativity, astrophysics
  and relativistic field theories. Proceedings, 12th Marcel Grossmann Meeting
  on General Relativity, Paris, France, July 12-18, 2009. Vol. 1-3}},
  pp.~2351--2355.
\newblock
2009.
\newblock

\bibitem{Bern:2018jmv}
Z.~Bern, J.~J. Carrasco, W.-M. Chen, A.~Edison, H.~Johansson,
  J.~Parra-Martinez, R.~Roiban, and M.~Zeng, ``{Ultraviolet Properties of
  $\mathcal N = 8$ Supergravity at Five Loops},''
  \href{http://dx.doi.org/10.1103/PhysRevD.98.086021}{{\em Phys. Rev.}
  {\bfseries D98} no.~8, (2018) 086021},
\href{http://arxiv.org/abs/1804.09311}{{\ttfamily arXiv:1804.09311 [hep-th]}}.

\bibitem{Green:2008uj}
M.~B. Green, J.~G. Russo, and P.~Vanhove, ``{Low energy expansion of the
  four-particle genus-one amplitude in type II superstring theory},''
  \href{http://dx.doi.org/10.1088/1126-6708/2008/02/020}{{\em JHEP} {\bfseries
  02} (2008) 020},
\href{http://arxiv.org/abs/0801.0322}{{\ttfamily arXiv:0801.0322 [hep-th]}}.

\bibitem{Green:2010sp}
M.~B. Green, J.~G. Russo, and P.~Vanhove, ``{String theory dualities and
  supergravity divergences},''
  \href{http://dx.doi.org/10.1007/JHEP06(2010)075}{{\em JHEP} {\bfseries 06}
  (2010) 075},
\href{http://arxiv.org/abs/1002.3805}{{\ttfamily arXiv:1002.3805 [hep-th]}}.

\bibitem{DHoker:2014oxd}
E.~D'Hoker, M.~B. Green, B.~Pioline, and R.~Russo, ``{Matching the $D^{6}R^{4}$
  interaction at two-loops},''
  \href{http://dx.doi.org/10.1007/JHEP01(2015)031}{{\em JHEP} {\bfseries 01}
  (2015) 031},
\href{http://arxiv.org/abs/1405.6226}{{\ttfamily arXiv:1405.6226 [hep-th]}}.

\bibitem{DHoker:2015gmr}
E.~D'Hoker, M.~B. Green, and P.~Vanhove, ``{On the modular structure of the
  genus-one Type II superstring low energy expansion},''
  \href{http://dx.doi.org/10.1007/JHEP08(2015)041}{{\em JHEP} {\bfseries 08}
  (2015) 041},
\href{http://arxiv.org/abs/1502.06698}{{\ttfamily arXiv:1502.06698 [hep-th]}}.

\bibitem{Virasoro:1969me}
M.~A. Virasoro, ``{Alternative constructions of crossing-symmetric amplitudes
  with regge behavior},''
\href{http://dx.doi.org/10.1103/PhysRev.177.2309}{{\em Phys. Rev.} {\bfseries
  177} (1969) 2309--2311}.

\bibitem{Green:1981yb}
M.~B. Green and J.~H. Schwarz, ``{Supersymmetrical String Theories},''
\href{http://dx.doi.org/10.1016/0370-2693(82)91110-8}{{\em Phys. Lett.}
  {\bfseries 109B} (1982) 444--448}.

\bibitem{DHoker:2001kkt}
E.~D'Hoker and D.~H. Phong, ``{Two loop superstrings. 1. Main formulas},''
  \href{http://dx.doi.org/10.1016/S0370-2693(02)01255-8}{{\em Phys. Lett.}
  {\bfseries B529} (2002) 241--255},
\href{http://arxiv.org/abs/hep-th/0110247}{{\ttfamily arXiv:hep-th/0110247
  [hep-th]}}.

\bibitem{DHoker:2005vch}
E.~D'Hoker and D.~H. Phong, ``{Two-loop superstrings VI: Non-renormalization
  theorems and the 4-point function},''
  \href{http://dx.doi.org/10.1016/j.nuclphysb.2005.02.043}{{\em Nucl. Phys.}
  {\bfseries B715} (2005) 3--90},
\href{http://arxiv.org/abs/hep-th/0501197}{{\ttfamily arXiv:hep-th/0501197
  [hep-th]}}.

\bibitem{Berkovits:2005ng}
N.~Berkovits and C.~R. Mafra, ``{Equivalence of two-loop superstring amplitudes
  in the pure spinor and RNS formalisms},''
  \href{http://dx.doi.org/10.1103/PhysRevLett.96.011602}{{\em Phys. Rev. Lett.}
  {\bfseries 96} (2006) 011602},
\href{http://arxiv.org/abs/hep-th/0509234}{{\ttfamily arXiv:hep-th/0509234
  [hep-th]}}.

\bibitem{DHoker:2017pvk}
E.~D'Hoker, M.~B. Green, and B.~Pioline, ``{Higher genus modular graph
  functions, string invariants, and their exact asymptotics},''
\href{http://arxiv.org/abs/1712.06135}{{\ttfamily arXiv:1712.06135 [hep-th]}}.

\bibitem{Penedones:2010ue}
J.~Penedones, ``{Writing CFT correlation functions as AdS scattering
  amplitudes},'' \href{http://dx.doi.org/10.1007/JHEP03(2011)025}{{\em JHEP}
  {\bfseries 03} (2011) 025},
\href{http://arxiv.org/abs/1011.1485}{{\ttfamily arXiv:1011.1485 [hep-th]}}.

\bibitem{Fitzpatrick:2011ia}
A.~L. Fitzpatrick, J.~Kaplan, J.~Penedones, S.~Raju, and B.~C. van Rees, ``{A
  Natural Language for AdS/CFT Correlators},''
  \href{http://dx.doi.org/10.1007/JHEP11(2011)095}{{\em JHEP} {\bfseries 11}
  (2011) 095},
\href{http://arxiv.org/abs/1107.1499}{{\ttfamily arXiv:1107.1499 [hep-th]}}.

\bibitem{Paulos:2011ie}
M.~F. Paulos, ``{Towards Feynman rules for Mellin amplitudes},''
  \href{http://dx.doi.org/10.1007/JHEP10(2011)074}{{\em JHEP} {\bfseries 10}
  (2011) 074},
\href{http://arxiv.org/abs/1107.1504}{{\ttfamily arXiv:1107.1504 [hep-th]}}.

\bibitem{Heemskerk:2009pn}
I.~Heemskerk, J.~Penedones, J.~Polchinski, and J.~Sully, ``{Holography from
  Conformal Field Theory},''
  \href{http://dx.doi.org/10.1088/1126-6708/2009/10/079}{{\em JHEP} {\bfseries
  10} (2009) 079},
\href{http://arxiv.org/abs/0907.0151}{{\ttfamily arXiv:0907.0151 [hep-th]}}.

\bibitem{Alday:2016njk}
L.~F. Alday, ``{Large Spin Perturbation Theory for Conformal Field Theories},''
  \href{http://dx.doi.org/10.1103/PhysRevLett.119.111601}{{\em Phys. Rev.
  Lett.} {\bfseries 119} no.~11, (2017) 111601},
\href{http://arxiv.org/abs/1611.01500}{{\ttfamily arXiv:1611.01500 [hep-th]}}.

\bibitem{Aharony:2016dwx}
O.~Aharony, L.~F. Alday, A.~Bissi, and E.~Perlmutter, ``{Loops in AdS from
  Conformal Field Theory},''
  \href{http://dx.doi.org/10.1007/JHEP07(2017)036}{{\em JHEP} {\bfseries 07}
  (2017) 036},
\href{http://arxiv.org/abs/1612.03891}{{\ttfamily arXiv:1612.03891 [hep-th]}}.

\bibitem{Alday:2017xua}
L.~F. Alday and A.~Bissi, ``{Loop Corrections to Supergravity on $AdS_5 \times
  S^5$},'' \href{http://dx.doi.org/10.1103/PhysRevLett.119.171601}{{\em Phys.
  Rev. Lett.} {\bfseries 119} no.~17, (2017) 171601},
\href{http://arxiv.org/abs/1706.02388}{{\ttfamily arXiv:1706.02388 [hep-th]}}.

\bibitem{Aprile:2017bgs}
F.~Aprile, J.~M. Drummond, P.~Heslop, and H.~Paul, ``{Quantum Gravity from
  Conformal Field Theory},''
  \href{http://dx.doi.org/10.1007/JHEP01(2018)035}{{\em JHEP} {\bfseries 01}
  (2018) 035},
\href{http://arxiv.org/abs/1706.02822}{{\ttfamily arXiv:1706.02822 [hep-th]}}.

\bibitem{Alday:2018pdi}
L.~F. Alday, A.~Bissi, and E.~Perlmutter, ``{Genus-One String Amplitudes from
  Conformal Field Theory},''
\href{http://arxiv.org/abs/1809.10670}{{\ttfamily arXiv:1809.10670 [hep-th]}}.

\bibitem{Caron-Huot:2017vep}
S.~Caron-Huot, ``{Analyticity in Spin in Conformal Theories},''
  \href{http://dx.doi.org/10.1007/JHEP09(2017)078}{{\em JHEP} {\bfseries 09}
  (2017) 078},
\href{http://arxiv.org/abs/1703.00278}{{\ttfamily arXiv:1703.00278 [hep-th]}}.

\bibitem{Beem:2016wfs}
C.~Beem, L.~Rastelli, and B.~C. van Rees, ``{More ${\mathcal N}=4$
  superconformal bootstrap},''
  \href{http://dx.doi.org/10.1103/PhysRevD.96.046014}{{\em Phys. Rev.}
  {\bfseries D96} no.~4, (2017) 046014},
\href{http://arxiv.org/abs/1612.02363}{{\ttfamily arXiv:1612.02363 [hep-th]}}.

\bibitem{Eden:2000bk}
B.~Eden, A.~C. Petkou, C.~Schubert, and E.~Sokatchev, ``{Partial
  nonrenormalization of the stress tensor four point function in N=4 SYM and
  AdS / CFT},'' \href{http://dx.doi.org/10.1016/S0550-3213(01)00151-1}{{\em
  Nucl. Phys.} {\bfseries B607} (2001) 191--212},
\href{http://arxiv.org/abs/hep-th/0009106}{{\ttfamily arXiv:hep-th/0009106
  [hep-th]}}.

\bibitem{Nirschl:2004pa}
M.~Nirschl and H.~Osborn, ``{Superconformal Ward identities and their
  solution},'' \href{http://dx.doi.org/10.1016/j.nuclphysb.2005.01.013}{{\em
  Nucl. Phys.} {\bfseries B711} (2005) 409--479},
\href{http://arxiv.org/abs/hep-th/0407060}{{\ttfamily arXiv:hep-th/0407060
  [hep-th]}}.

\bibitem{Rastelli:2017udc}
L.~Rastelli and X.~Zhou, ``{How to Succeed at Holographic Correlators Without
  Really Trying},'' \href{http://dx.doi.org/10.1007/JHEP04(2018)014}{{\em JHEP}
  {\bfseries 04} (2018) 014},
\href{http://arxiv.org/abs/1710.05923}{{\ttfamily arXiv:1710.05923 [hep-th]}}.

\bibitem{Alday:2014tsa}
L.~F. Alday, A.~Bissi, and T.~Lukowski, ``{Lessons from crossing symmetry at
  large N},'' \href{http://dx.doi.org/10.1007/JHEP06(2015)074}{{\em JHEP}
  {\bfseries 06} (2015) 074},
\href{http://arxiv.org/abs/1410.4717}{{\ttfamily arXiv:1410.4717 [hep-th]}}.

\bibitem{Alday:2017vkk}
L.~F. Alday and S.~Caron-Huot, ``{Gravitational S-matrix from CFT dispersion
  relations},'' \href{http://dx.doi.org/10.1007/JHEP12(2018)017}{{\em JHEP}
  {\bfseries 12} (2018) 017},
\href{http://arxiv.org/abs/1711.02031}{{\ttfamily arXiv:1711.02031 [hep-th]}}.

\bibitem{Goncalves:2014ffa}
V.~Goncalves, ``{Four point function of $\mathcal{N}=4$ stress-tensor multiplet
  at strong coupling},'' \href{http://dx.doi.org/10.1007/JHEP04(2015)150}{{\em
  JHEP} {\bfseries 04} (2015) 150},
\href{http://arxiv.org/abs/1411.1675}{{\ttfamily arXiv:1411.1675 [hep-th]}}.

\bibitem{Chester:2018dga}
S.~M. Chester and E.~Perlmutter, ``{M-Theory Reconstruction from (2,0) CFT and
  the Chiral Algebra Conjecture},''
  \href{http://dx.doi.org/10.1007/JHEP08(2018)116}{{\em JHEP} {\bfseries 08}
  (2018) 116},
\href{http://arxiv.org/abs/1805.00892}{{\ttfamily arXiv:1805.00892 [hep-th]}}.

\bibitem{Gopakumar:2016wkt}
R.~Gopakumar, A.~Kaviraj, K.~Sen, and A.~Sinha, ``{Conformal Bootstrap in
  Mellin Space},'' \href{http://dx.doi.org/10.1103/PhysRevLett.118.081601}{{\em
  Phys. Rev. Lett.} {\bfseries 118} no.~8, (2017) 081601},
\href{http://arxiv.org/abs/1609.00572}{{\ttfamily arXiv:1609.00572 [hep-th]}}.

\bibitem{Gopakumar:2016cpb}
R.~Gopakumar, A.~Kaviraj, K.~Sen, and A.~Sinha, ``{A Mellin space approach to
  the conformal bootstrap},''
  \href{http://dx.doi.org/10.1007/JHEP05(2017)027}{{\em JHEP} {\bfseries 05}
  (2017) 027},
\href{http://arxiv.org/abs/1611.08407}{{\ttfamily arXiv:1611.08407 [hep-th]}}.

\bibitem{Gopakumar:2018xqi}
R.~Gopakumar and A.~Sinha, ``{On the Polyakov-Mellin bootstrap},''
  \href{http://dx.doi.org/10.1007/JHEP12(2018)040}{{\em JHEP} {\bfseries 12}
  (2018) 040},
\href{http://arxiv.org/abs/1809.10975}{{\ttfamily arXiv:1809.10975 [hep-th]}}.

\bibitem{Ghosh:2018bgd}
K.~Ghosh, ``{Polyakov-Mellin Bootstrap for AdS loops},''
\href{http://arxiv.org/abs/1811.00504}{{\ttfamily arXiv:1811.00504 [hep-th]}}.

\bibitem{Mazac:2018ycv}
D.~Mazac and M.~F. Paulos, ``{The Analytic Functional Bootstrap II: Natural
  Bases for the Crossing Equation},''
\href{http://arxiv.org/abs/1811.10646}{{\ttfamily arXiv:1811.10646 [hep-th]}}.

\bibitem{Mazac:2018qmi}
D.~Mazac, ``{A Crossing-Symmetric OPE Inversion Formula},''
\href{http://arxiv.org/abs/1812.02254}{{\ttfamily arXiv:1812.02254 [hep-th]}}.

\bibitem{Cardona:2018nnk}
C.~Cardona, ``{OPE inversion in Mellin space},''
\href{http://arxiv.org/abs/1803.05086}{{\ttfamily arXiv:1803.05086 [hep-th]}}.

\bibitem{Cardona:2018dov}
C.~Cardona and K.~Sen, ``{Anomalous dimensions at finite conformal spin from
  OPE inversion},'' \href{http://dx.doi.org/10.1007/JHEP11(2018)052}{{\em JHEP}
  {\bfseries 11} (2018) 052},
\href{http://arxiv.org/abs/1806.10919}{{\ttfamily arXiv:1806.10919 [hep-th]}}.

\bibitem{Zhou:2018sfz}
X.~Zhou, ``{Recursion Relations in Witten Diagrams and Conformal Partial
  Waves},''
\href{http://arxiv.org/abs/1812.01006}{{\ttfamily arXiv:1812.01006 [hep-th]}}.

\bibitem{Chester:2018aca}
S.~M. Chester, S.~S. Pufu, and X.~Yin, ``{The M-Theory S-Matrix From ABJM:
  Beyond 11D Supergravity},''
  \href{http://dx.doi.org/10.1007/JHEP08(2018)115}{{\em JHEP} {\bfseries 08}
  (2018) 115},
\href{http://arxiv.org/abs/1804.00949}{{\ttfamily arXiv:1804.00949 [hep-th]}}.

\bibitem{DHoker:2015wxz}
E.~D'Hoker, M.~B. Green, Ö.~Gürdogan, and P.~Vanhove, ``{Modular Graph
  Functions},'' \href{http://dx.doi.org/10.4310/CNTP.2017.v11.n1.a4}{{\em
  Commun. Num. Theor. Phys.} {\bfseries 11} (2017) 165--218},
\href{http://arxiv.org/abs/1512.06779}{{\ttfamily arXiv:1512.06779 [hep-th]}}.

\bibitem{Vanhove:2018elu}
P.~Vanhove and F.~Zerbini, ``{Closed string amplitudes from single-valued
  correlation functions},''
\href{http://arxiv.org/abs/1812.03018}{{\ttfamily arXiv:1812.03018 [hep-th]}}.

\bibitem{Broedel:2018izr}
J.~Broedel, O.~Schlotterer, and F.~Zerbini, ``{From elliptic multiple zeta
  values to modular graph functions: open and closed strings at one loop},''
\href{http://arxiv.org/abs/1803.00527}{{\ttfamily arXiv:1803.00527 [hep-th]}}.

\bibitem{Zerbini:2018hgs}
F.~Zerbini, ``{Modular and holomorphic graph function from superstring
  amplitudes},'' in {\em {KMPB Conference: Elliptic Integrals, Elliptic
  Functions and Modular Forms in Quantum Field Theory Zeuthen, Germany, October
  23-26, 2017}}.
\newblock 2018.
\newblock
\href{http://arxiv.org/abs/1807.04506}{{\ttfamily arXiv:1807.04506 [math-ph]}}.
\newblock

\bibitem{DHoker:2016mwo}
E.~D'Hoker and M.~B. Green, ``{Identities between Modular Graph Forms},'' {\em
  J. Number Theor.} {\bfseries 189} (2018) 25--88,
\href{http://arxiv.org/abs/1603.00839}{{\ttfamily arXiv:1603.00839 [hep-th]}}.

\bibitem{Basu:2016kli}
A.~Basu, ``{Proving relations between modular graph functions},''
  \href{http://dx.doi.org/10.1088/0264-9381/33/23/235011}{{\em Class. Quant.
  Grav.} {\bfseries 33} no.~23, (2016) 235011},
\href{http://arxiv.org/abs/1606.07084}{{\ttfamily arXiv:1606.07084 [hep-th]}}.

\bibitem{Brown:2017qwo}
F.~Brown, ``{A class of non-holomorphic modular forms I},''
\newblock 2017.
\newblock
\href{http://arxiv.org/abs/1707.01230}{{\ttfamily arXiv:1707.01230 [math.NT]}}.
\newblock

\bibitem{Caron-Huot:2018kta}
S.~Caron-Huot and A.-K. Trinh, ``{All Tree-Level Correlators in
  AdS${}_5\times$S${}_5$ Supergravity: Hidden Ten-Dimensional Conformal
  Symmetry},''
\href{http://arxiv.org/abs/1809.09173}{{\ttfamily arXiv:1809.09173 [hep-th]}}.

\end{thebibliography}\endgroup

\end{document}